\documentclass[preprint,useAMS,usenatbib]{mn2e}
\usepackage{amsmath}
\usepackage{natbib}
\usepackage{graphicx}
\usepackage{graphics}
\usepackage{subfigure}
\usepackage{amssymb}
\usepackage{amsfonts}
\usepackage{verbatim}
\usepackage{times}
\usepackage[usenames]{color}
\usepackage{longtable}
\usepackage{supertabular}
\usepackage{array}
\usepackage{chngpage}
\usepackage{ragged2e}
\usepackage{array}
\usepackage{rotating}
\usepackage{lscape}
\usepackage{multicol}

\renewcommand{\raggedright}{\leftskip=0pt \rightskip=0pt plus 0cm}
\newcommand       \kpc          {\,{\rm kpc}}

\newcommand       \simali       {{\sim}\,}
\newcommand       \magni        {\,{\rm mag}}
\newcommand       \magniAL    {{\rm mag}}
\newcommand     \gtsim  {\lower.5ex\hbox{$\buildrel > \over \sim$}}
\newcommand     \ltsim  {\lower.5ex\hbox{$\buildrel < \over \sim$}}
\newcommand     \simgt  {\lower.5ex\hbox{$\buildrel > \over \sim$}}
\newcommand     \simlt  {\lower.5ex\hbox{$\buildrel < \over \sim$}}
\newcommand{\PreserveBackslash}[1]{\let\temp=\\#1\let\\=\temp}
\newcolumntype{C}[1]{>{\PreserveBackslash\centering}p{#1}}
\newcolumntype{R}[1]{>{\PreserveBackslash\raggedleft}p{#1}}
\newcolumntype{L}[1]{>{\PreserveBackslash\raggedright}p{#1}}

\newcommand       \Angstrom     {\,{\rm \AA}}

\newcommand       \cm           {\,{\rm cm}}

\newcommand       \pc           {\,{\rm pc}}

\newcommand       \keV           {\,{\rm keV}}
\newcommand       \nH           {n_{\rm H}}
\newcommand       \NH           {N_{\rm H}}
\newcommand       \rmH          {\,{\rm H}}
\newcommand       \HH          {{\rm H}}
\newcommand       \rmHAL      {{\rm H}}

\newcommand       \Msun       {\,{M_{\rm \bigodot}}}
\newcommand{\AV}{A_V}
\newcommand{\Av}{A_V}
\newcommand{\RV}{R_V}

\newcommand       \Mgas       {M_{\rm gas}}
\newcommand       \Mdust      {M_{\rm dust}}
\newcommand       \muH        {\mu_{\rm H}}
\newcommand       \kappaV    {\kappa_{\rm ext}(V)}
\newcommand       \mnras        {MNRAS}
\newcommand       \apj          {ApJ}
\newcommand       \apjs         {ApJS}
\newcommand       \apjl         {ApJ}
\newcommand       \aj           {AJ}
\newcommand       \aap          {A\&A}
\newcommand       \aaps         {A\&AS}
\newcommand       \araa         {ARA\&A}
\newcommand       \pasj         {PASJ}
\newcommand       \memsai       {MmSAI}
\newcommand       \apss         {AP\&SS}
\newcommand       \sovast       {SvA}

\title[The Galactic $N_{\rm H}$--$\AV$ Relation]
        {The gas-to-extinction ratio
         and the gas distribution in the Galaxy}
\author[Zhu, Tian, Li \& Zhang]
       {Hui~Zhu$^{1,2,3}$,
        Wenwu~Tian$^{1,4,5}$\thanks{E-mail: tww@bao.ac.cn},
        Aigen~Li$^{3}$
        and Mengfei~Zhang$^{1,4}$\\
$^1$Key Laboratory of Optical Astronomy,
    National Astronomical Observatories,
    Chinese Academy of Sciences,
    Beijing 100012, China\\
$^2$Harvard-Smithsonian Center for Astrophysics,
    60 Garden Street, Cambridge, MA 02138, USA\\
$^3$Department of Physics and Astronomy,
    University of Missouri,
    Columbia, MO 65211, USA\\
$^4$School of Astronomy and Space Science, University of Chinese Academy of Sciences,
    Beijing 100049, China\\
$^5$Department of Physics and Astronomy,
    University of Calgary,
    Calgary, Alberta T2N 1N4, Canada\\
    }

\date{Accepted 000. Received 000; in original form 000}

\begin{document}

\maketitle

\begin{abstract}
We investigate the relation between
the optical extinction ($\AV$)
and the hydrogen column density ($\NH$)
determined from X-ray observations
of a large sample of Galactic sightlines
toward 35 supernova remnants,
6 planetary nebulae, and 70 X-ray binaries
for which $\NH$ was determined in the literature
with solar abundances.
% of Ander \& Grevesse (1989).
%
We derive an average ratio of
$\langle\NH/\AV\rangle=\left(2.08\pm0.02\right)\times10^{21}\rmH\cm^{-2}\magni^{-1}$
for the whole Galaxy.
We find no correlation between
$\langle\NH/\AV\rangle$ and
the number density of hydrogen,
the distance away from the Galactic centre,
and the distance above or below the Galactic plane.
The $\langle\NH/\AV\rangle$ ratio
is generally invariant across the Galaxy,
with $\langle\NH/\AV\rangle=\left(2.04\pm0.05\right)\times10^{21}\rmH\cm^{-2}\magni^{-1}$
for the 1st and 4th Galactic quadrants and
$\langle\NH/\AV\rangle=\left(2.09\pm0.03\right)\times10^{21}\rmH\cm^{-2}\magni^{-1}$
for the 2nd and 3rd Galactic quadrants.
We also explore the distribution of
hydrogen in the Galaxy by enlarging our sample
with additional 74 supernova remnants
for which both $\NH$ and distances are known.
We find that, between the Galactic radius of 2\,kpc to 10\,kpc,
the vertical distribution of hydrogen can be roughly
described by a Gaussian function with a scale height of
$h=75.5\pm12.4\pc$
and a mid-plane density of
$n_{\rm H}(0)=1.11\pm0.15\cm^{-3}$,
corresponding to a total gas surface density
of $\sum_{\rm gas}\simali7.0\Msun\pc^{-2}$.
We also compile $\NH$
from 19 supernova remnants and
29 X-ray binaries for which $\NH$
was determined with subsolar abundances.
% of Wilms et al. (2000).
We obtain $\langle\NH/\AV\rangle=\left(2.47\pm0.04\right)\times10^{21}\rmH\cm^{-2}\magni^{-1}$
which exceeds that derived with
solar abundances by $\simali$20\%.
We suggest that in future studies one
may simply scale $\NH$ derived from
subsolar abundances
by a factor of $\simali$1.2
when converting to $\NH$ of solar abundances.
\end{abstract}


\section{Introduction}
Interstellar extinction results from the absorption
and scattering of starlight by interstellar dust grains.
The knowledge of the interstellar extinction ($A_\lambda$)
at wavelength $\lambda$ per hydrogen column ($\NH$),
particularly in the ultraviolet (UV), visible, and near infrared,
is crucial for correcting for the interstellar obscuration
and restoring the true luminosity of astronomical objects,
and for determining the dust-to-gas mass ratio
and the photoelectric heating rates of interstellar gas.
So far, the commonly adopted hydrogen-to-extinction ratio
is that of Bohlin et al.\ (1978),
$\NH/E(B-V) =  5.8\times10^{21}\rmH\cm^{-2}\magni^{-1}$,
where $E(B-V)\equiv A_B-\AV$, the colour excess,
is the difference between the $B$ band extinction ($A_B$)
and the $V$ band extinction ($\AV$).
The colour excess $E(B-V)$ relates to $\AV$
through $\RV\equiv\AV/E(B-V)$,
the total-to-selective extinction ratio.
%

%%%%% Table 1 %%%%%
\begin{table*}
\caption{$\NH/\AV$ Ratios Determined
         from the HI and H$_2$ UV Absorption Spectra.}
%\begin{minipage}[b]{1.0\textwidth}
\begin{center}
\begin{tabular}{lcccr}
\hline
\hline
Hydrogen &  $\NH/\AV$ &  Sample &  Reference       &   Comments \\
Type     & ($10^{21}\cm^{-2}\magni^{-1}\rmH$) & Size & & \\
\hline
HI+H$_2$ & 2.15$\pm$0.14 & 17 & Rachford et al.\ (2009)
         & {\it FUSE} observations of stars with $E(B-V)\simlt0.85\magni$ \\
HI & 1.59$\pm$0.09 & 393 & Diplas \& Savage (1994)
         & {\it IUE} observations of stars with $E(B-V)\simlt0.6\magni$ \\
HI+H$_2$ & 1.9 & 12 &  Whittet (1981)
         & {\it Copernicus} observations of stars
           with $E(B-V)\simlt0.6\magni$\\
HI+H$_2$ & 1.87 & 75 & Bohlin et al.\ (1978)
         & {\it Copernicus} observation of stars
           with $E(B-V)\simlt0.5\magni$ \\
HI+H$_2$ & 2.0  & 95 & Jenkins \& Savage (1974)
         & {\it OAO-2} observations of stars
           with $E(B-V)\simlt0.4\magni$\\
\hline
\end{tabular}
\end{center}
\end{table*}
%%%%% Table 1 %%%%%

%%%%% Table 2 %%%%%
\begin{table*}
\caption{$\NH/\AV$ Ratios Determined
         from the HI 21\,cm Emission and/or CO emission.}
%\begin{minipage}[b]{1.0\textwidth}
\begin{center}
\begin{tabular}{lcccr}
\hline
\hline
Hydrogen &  $\NH/\AV$ &  Sample &  Reference       &   Comments \\
Type     & ($10^{21}\cm^{-2}\magni^{-1}\rmH$) & Size & & \\
\hline
HI & 2.68 & -- &  Liszt (2014)
          & Galactic latitudes $|b|\gtsim20^{\circ}$
            and $E(B-V)\simlt0.1\magni$ \\
HI & 1.56 & 76 &  Heiles (1976) & stars and globular clusters
                  with $E(B-V)\simlt0.6\magni$ \\
HI & 1.6 & 81 &  Knapp \& Kerr (1974)
         & globular clusters with $E(B-V)\ltsim1.3\magni$ \\
HI & $\simali$1.3--1.6 & 40 &   Sturch (1969)
         & RR Lyrae stars
           with $E(B-V)\simlt0.4\magni$ \\
HI+H$_2$ & 2.4 & -- & Chen et al.\ (2015)
         & Galactic latitudes $|b|\gtsim10^{\circ}$
         and $E(B-V)\simlt1.0\magni$\\
\hline
\end{tabular}
\end{center}
\end{table*}
%%%%% Table 2 %%%%%

%%%% Table 3 %%%
\begin{table*}
{\tiny
\caption{$\NH/\AV$ Ratios Previously Determined
         from Interstellar X-Ray Absorption Data.}
\begin{center}
\begin{tabular}{ccccccccl}
\hline
\hline
$\NH/\AV$ & Sample
          & Interstellar Abundance & O/H & Mg/H & Si/H & Fe/H
          & Reference & Comments \\
($10^{21}\rmH\cm^{-2}\magni^{-1}$) & Size
          & Standard & (ppm) & (ppm) & (ppm) & (ppm)
          & &\\
\hline
2.87$\pm$0.12 & 17
    & Wilms et al.\ (2000) & 4.90 & 2.51 & 1.86 & 2.69
    & Foight et al.\ (2016)
    & SNRs with $E(B-V)\simlt10\magni$ \\
2.08$\pm$0.30 & 23
    & Anders \& Grevesse (1989) & 8.51 & 3.80 & 3.55 & 4.68
    & Valencic \& Smith (2015)
    & XBs with $E(B-V)\simlt2.5\magni$ \\
2.2$^{+0.3}_{-0.4}$ & $>$100
    & Anders \& Grevesse (1989) & 8.51 &  3.80 & 3.55 & 4.68
    & Watson (2011)
    %& obtained by fitting the $\NH/\Av$ upper-bound of gamma-ray bursts\\
    & $\gamma$-ray bursts\\
2.21$\pm$0.09 & 22
    &  Anders \& Grevesse (1989) & 8.51 & 3.80 & 3.55 & 4.68
    & Guver \& Ozel (2009)
    & SNRs with $E(B-V)\ltsim10\magni$ \\
$\simali$1 & 20
    & Anders \& Ebihara (1982) & 7.39 & 3.95 & 3.68 & 3.31
    & Vuong et al.\ (2003)
    & SF regions with distances$\simlt$0.5\,kpc \\
1.79$\pm$0.03 & 26
    & Anders \& Ebihara (1982) & 7.39 & 3.95 & 3.68 & 3.31
    & Predehl \& Schmitt (1995)
    & SNRs and XBs with $E(B-V)\simlt6.1\magni$ \\
2.19$\pm$0.52 &  5
    &  - & - & - & - & -
    & Ryter et al.\ (1975)
    & SNRs with $E(B-V)\ltsim10\magni$ \\
2.22$\pm$0.14 & 7
    &  - & - & - & - & -
    & Gorenstein (1975)
    & SNRs with $E(B-V)\simlt10\magni$ \\
1.85 & 4
     & - & - & - & - & -
     & Reina \& Tarenghi (1973)
     & XBs, extended sources
       with $E(B-V)$$\simlt$10$\magni$ \\
\hline
\end{tabular}
\end{center}
}
\end{table*}
%%%% Table 3 %%%

%%%% Figure 1 %%%%
\begin{figure}
\centerline{\includegraphics[width=0.49\textwidth, angle=0]{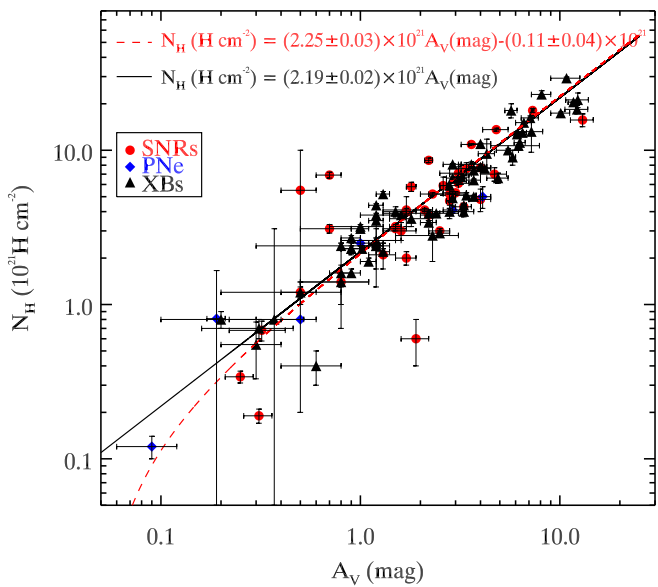}}
\caption{Hydrogen column densities $\NH$ against
        the optical extinction $\AV$
        for the AG89 sample.
        Red circles: SNRs, Blue rhombus: PNe,
         Black triangles: XBs.
        }
\label{fig1}
\end{figure}
%%%% Figure 1 %%%%

%%%% Table 4 %%%%
\begin{table*}
\caption{Optical extinction $\AV$
         and hydrogen column density $\NH$
         compiled from the literature
         (see Appendix for details).
         }
         {\scriptsize
\begin{minipage}[b]{1.0\textwidth}
\begin{center}
\begin{tabular}{@{}p{1.9cm}p{0.7cm}p{0.7cm}p{1.4cm}p{1.4cm}p{1.4cm}p{1.4cm}p{0.9cm}p{0.6cm}p{1.9cm}p{1.4cm}@{}}
\hline
\hline
Object & Gl$^a$  &   Gb$^a$
     & $\NH({\rm KBH})^b$
     & $\NH({\rm DL})^c$
     & ${\NH}({\rm AG89})^d$
     & ${\NH}({\rm W00})^e$
     & $\AV$
     & $D^f$
     & ${\NH}({\rm AG89})$/$\AV$
     & ${\NH}({\rm W00})$/$\AV$\\
Name & (degree) & (degree)
     & ($10^{21}\rmH\cm^{-2}$)
     & ($10^{21}\rmH\cm^{-2}$)
     & ($10^{21}\rmH\cm^{-2}$)
     & ($10^{21}\rmH\cm^{-2}$)
     & ($\magni$)
     & (kpc)
     & {\tiny ($10^{21}\rmH\cm^{-2}{\magni}^{-1}$)}
     & {\tiny ($10^{21}\rmH\cm^{-2}{\magni}^{-1}$)}\\
\hline
&            &            &            &            &             &    &   & & &     \\
\multicolumn{ 11}{c}{Supernova remnants} \\
&            &            &            &            &             &    &   & & &      \\
G4.5+6.8 north      &        4.5 &        6.8 &      2.12  &      2.38  &        5.1$\pm$0.3 & 5.4$\pm$0.1 &      3.0$\pm$0.4 & 5.1 & 1.89$\pm$0.27 & 1.93$\pm$0.28\\
G6.4-0.1 central    &        6.4 &       -0.1 &     11.40  &     14.40  &        4.3$\pm$0.1 & 6.3$\pm$1.9 &      3.3$\pm$0.3 & 1.9 & 1.30$\pm$0.12 & 1.91$\pm$0.60\\
G6.4-0.1 northeast  &        6.4 &       -0.1 &     11.40  &     14.40  &        7.7$\pm$0.2 &             &      4.0$\pm$0.5 & 2.0 & 1.93$\pm$0.25 & \\
G11.2-0.3 southeast &       11.2 &        0.3 &     12.70  &     17.50  &       15.7$\pm$1.5 &   &   13.0$\pm$1.7 & 4.4 & 1.21$\pm$0.20 & \\
G13.3-1.3           &       13.3 &       -1.3 &      8.05  &      8.11  &        5.5$\pm$4.5 &   &    0.5$\pm$0.1 & 3.0 & 11.00$\pm$9.26 & \\
G39.7-2.0 east jet  &       39.7 &       -2.0 &      6.49  &      7.63  &        5.2$\pm$0.1 &   &    2.3$\pm$0.3 & 5.5 & 2.26$\pm$0.30 &\\
G53.6-2.2 central   &       53.6 &       -2.2 &      5.48  &      6.69  &        6.1$\pm$0.1 &   &   3.1$\pm$0.4 & 2.3 & 1.97$\pm$0.26 &\\
G53.6-2.2 southwest &       53.6 &       -2.2 &      5.48  &      6.69  &        4.8$\pm$0.8 & 6.7$\pm$0.3 &      4.0$\pm$0.5 & 2.3 & 1.20$\pm$0.25 & 1.68$\pm$0.22\\
G54.1+0.3           &       54.1 &        0.3 &     11.90  &     14.30  &       18.1$\pm$0.5 &  25.5$\pm$0.4 &     7.3$\pm$0.1 & 6.2 & 2.48$\pm$0.08 & 3.49$\pm$0.07\\
G67.7+1.8 north     &       67.7 &        1.8 &      9.42  &     10.50  &        4.1$\pm$0.9 &    &   1.7$\pm$0.4 &  & 2.41$\pm$0.78 & \\
G69.0+2.7 central   &         69 &        2.7 &      6.59  &      7.92  &        3.0$\pm$0.1 &  4.5$\pm$0.2  &   2.5$\pm$0.3 & 1.5 & 1.20$\pm$0.15 & 1.80$\pm$0.23\\
G74.0-8.5           &       74.0 &       -8.5 &      1.60  &      1.68  &      0.34$\pm$0.03 &    & 0.25$\pm$0.06 & 0.54 & 1.36$\pm$0.25 & \\
G78.2+2.1           &       78.2 &        2.1 &     12.10  &     13.80  &        7.6$\pm$0.7 &    &   3.4$\pm$0.6 & 2.0 & 2.24$\pm$0.44 & \\
G82.2+5.3           &       82.2 &        5.3 &      6.87  &      8.30  &        4.7$\pm$0.5 &    &   2.8$\pm$0.1 & 2.0 & 1.68$\pm$0.19 & \\
G85.9-0.6           &       85.9 &       -0.6 &      7.59  &      9.45  &        6.9$\pm$0.3 &    &   0.7$\pm$0.1 & 5.0 & 9.86$\pm$1.47 & \\
G89.0+4.7 northwest &       89.0 &        4.7 &      6.83  &      7.64  &        3.0$\pm$0.4 &    &   1.6$\pm$0.3 & 1.7 & 1.88$\pm$0.43 & \\
G109.1-1.0          &      109.1 &       -1.0 &      6.39  &      7.50  &       7.1$\pm$0.7  &  9.0$\pm$1.4  &   3.1$\pm$0.6 & 3.2 & 2.29$\pm$0.50 & 2.90$\pm$0.72 \\
G111.7-2.1 north    &      111.7 &       -2.1 &      3.94  &      6.70  &      13.6$\pm$0.3  &  11.7$\pm$1.6 &     4.8$\pm$0.7 & 3.4 & 2.83$\pm$0.42 & 2.44$\pm$0.49 \\
G116.9+0.2          &      116.9 &        0.2 &      6.32  &      6.77  &        5.9$\pm$0.8 &  9.2$\pm$0.7  &     2.6$\pm$0.1 & 3.0 & 2.27$\pm$0.32 & 3.54$\pm$0.30 \\
G119.5+10.2         &      119.5 &       10.2 &      2.06  &      2.17  &        2.1$\pm$0.4 &  3.8$\pm$1.1  &     1.3$\pm$0.2 & 1.4 & 1.62$\pm$0.40 & 2.92$\pm$0.96 \\
G120.1+1.4          &      120.1 &        1.4 &      8.24  &      9.22  &        5.8$\pm$0.4 &  8.0$\pm$1.0  &     1.8$\pm$0.1 & 3.0 & 3.22$\pm$0.29 & 4.44$\pm$0.61 \\
G130.7+3.1          &      130.7 &        3.1 &      5.57  &      5.72  &        4.1$\pm$0.1 &  5.4$\pm$0.1  &     2.1$\pm$0.2 & 2.0 & 1.95$\pm$0.19 & 2.57$\pm$0.25\\
G132.7+1.3          &      132.7 &        1.3 &      7.98  &      9.00  &        8.3$\pm$0.3 &   &    2.2$\pm$0.1 & 2.5 & 3.91$\pm$0.22 & \\
G166.0+4.3          &      166.0 &        4.3 &      3.86  &      5.67  &        2.0$\pm$0.2 &   &    1.7$\pm$0.2 & 4.5 & 1.18$\pm$0.18 & \\
G180.0-1.7          &      180.0 &       -1.7 &      4.93  &      6.30  &        3.1$\pm$0.2 &   &    0.7$\pm$0.2 & 1.3 & 4.43$\pm$1.30 & \\
G184.6-5.8          &      184.6 &       -5.8 &      3.26  &      3.79  &        3.2$\pm$0.1 &  4.1$\pm$0.1  &     1.5$\pm$0.1 & 2.0 & 2.13$\pm$0.16 & 2.73$\pm$0.19\\
G189.1+3.0          &      189.1 &        3.0 &      4.85  &      6.09  &       5.2$\pm$0.5  &  6.0$\pm$0.5  &     2.8$\pm$0.6 & 1.5 & 1.86$\pm$0.44 & 2.14$\pm$0.49\\
G260.4-3.4          &      260.4 &       -3.4 &      7.45  &      9.46  &        3.1$\pm$0.1 &  3.5$\pm$1.5  &     1.5$\pm$0.2 & 2.2 & 2.07$\pm$0.28 & 2.33$\pm$1.05 \\
G263.9-3.3          &      263.9 &       -3.3 &      7.38  &      9.10  &      0.19$\pm$0.02 &  0.3$\pm$0.1  &   0.31$\pm$0.05 & 0.29 & 0.61$\pm$0.12 & 0.97$\pm$0.36\\
G292.0+1.8          &        292 &        1.8 &      5.52  &      6.21  &       3.8$\pm$0.3  &  5.0$\pm$0.4  &     2.2$\pm$0.2 & 6.0 & 1.73$\pm$0.21 & 2.27$\pm$0.28 \\
G296.1-0.5 north    &      296.1 &       -0.5 &     13.20  &     16.20  &       0.6$\pm$0.2  &   &    1.9$\pm$0.3 & 2.0 & 0.32$\pm$0.12 & \\
G296.5+10.0         &      296.5 &       10.0 &      1.13  &      1.27  &       1.2$\pm$0.1  &   &    0.5$\pm$0.1 & 2.1 & 2.40$\pm$0.52 & \\
G315.4-2.3southwest &     315.4 &       -2.3 &      7.19  &      9.06  &       4.0$\pm$0.2  &   &    1.7$\pm$0.2 & 2.3 & 2.35$\pm$0.30 & \\
G320.4-1.2          &      320.4 &       -1.2 &     14.80  &     14.70  &      10.9$\pm$0.1  & 11.5$\pm$0.3  &      3.6$\pm$0.4 & 5.2 & 3.03$\pm$0.34 & 3.19$\pm$0.36 \\
G327.6+14.6         &      327.6 &       14.7 &      0.75  &      0.68  &     0.68$\pm$0.10  & 1.6$\pm$0.1   &    0.32$\pm$0.10 & 2.2 & 2.13$\pm$0.73 & 5.00$\pm$1.59 \\
G332.4-0.4          &      332.4 &       -0.4 &     20.70  &     22.80  &       7.0$\pm$0.7  & 8.7$\pm$3.5   &      4.7$\pm$0.7 & 3.1 & 1.49$\pm$0.27 & 1.85$\pm$0.79 \\
G332.5-5.6          &      332.5 &       -5.6 &      2.47  &      2.84  &       1.4$\pm$0.4  &    &   0.8$\pm$0.3 & 3.0 & 1.75$\pm$0.83 & \\
&            &            &            &            &             &    &   & & &    \\
\hline
&            &            &            &            &             &    &   & & &    \\
\multicolumn{ 11}{c}{Planetary nebulae} \\
&            &            &            &            &             &    &   & & &   \\
G37.7-34.5          &       37.7 &      -34.5 &      0.50  &      0.47  &        0.81$\pm$0.85 &   &  0.19$\pm$0.02 & 1.1 & 4.26$\pm$4.50 & \\
G64.7+5.0           &       64.7 &        5.0 &      3.03  &      3.10  &        2.5$\pm$0.1 &    &   1.0$\pm$0.1 & 1.2 & 2.50$\pm$0.27 & \\
G84.9-3.4           &       84.9 &       -3.4 &      4.68  &      5.58  &        4.1$\pm$0.2 &   &    2.9$\pm$0.3 & 0.7 & 1.41$\pm$0.16 & \\
G94.0+27.4          &       94.0 &       27.4 &      0.34  &      0.41  &      0.12$\pm$0.02 &   &  0.09$\pm$0.03 & 1.0 & 1.33$\pm$0.50 & \\
G197.8+17.3         &      197.8 &       17.3 &      0.54  &      0.57  &        0.8$\pm$0.6 &   &    0.5$\pm$0.1 & 1.2 & 1.60$\pm$1.24 & \\
G331.7-1.0          &      331.7 &       -1.0 &     16.70  &     17.60  &        5.0$\pm$0.8 &   &    4.1$\pm$0.4 & 1.1 & 1.22$\pm$0.23 & \\
&            &            &            &            &             &    &   & & &   \\
\hline
&            &            &            &            &             &    &   & & &   \\
\multicolumn{ 11}{c}{X-ray Binaries} \\
&            &            &            &            &             &    &   & & &   \\
2E 0236.6+6101      &      135.7 &        1.1 &      7.57  &      9.02  &       5.9$\pm$0.2  &     &   2.8$\pm$0.3 & 2.4 & 2.11$\pm$0.24 & \\
2E 1118.7-6138      &      292.5 &       -0.9 &     12.00  &     14.10  &        8.1$\pm$0.1 &  13.2$\pm$0.3  &      2.9$\pm$0.3 & 5.0 & 2.79$\pm$0.29 & 4.55$\pm$0.48 \\
2S 0620-003         &      209.6 &       -6.5 &      3.66  &      4.10  &        2.4$\pm$1.1 &   &     1.2$\pm$0.1 & 1.2 & 2.00$\pm$0.93 & \\
2S 0921-63          &      281.8 &       -0.9 &      1.70  &      2.13  &        2.3$\pm$0.1 &   &   1.02$\pm$0.03 & 8.5 & 2.25$\pm$0.12 & \\
4U 0352+30          &      163.1 &      -17.1 &      7.09  &      9.39  &        2.5$\pm$1.1 &    &    1.2$\pm$0.1 & 1.0 & 2.08$\pm$0.19 & \\
4U 0538+26          &      181.4 &       -2.6 &      4.50  &      5.90  &        3.9$\pm$0.1 &  7.0$\pm$0.3   &      2.4$\pm$0.2 & 2.5 & 1.63$\pm$0.08 & 2.92$\pm$0.17 \\
4U 0614+09          &      200.9 &       -3.4 &      4.48  &      5.47  &       3.5$\pm$0.1  &  3.3$\pm$0.1   &      1.2$\pm$0.1 & & 2.92$\pm$0.26 & 2.75$\pm$0.24 \\
4U 0919-54          &      275.9 &       -3.8 &      6.16  &      7.30  &        3.1$\pm$0.1 &  3.0$\pm$0.1   &     1.0$\pm$0.1  & 4.8 & 3.10$\pm$0.33 & 3.00$\pm$0.32 \\
4U 1036-56          &      285.4 &        1.4 &      9.02  &     10.10  &        2.8$\pm$0.9 &   &    2.3$\pm$0.8  & 5.0 & 1.22$\pm$0.58 & \\
4U 1118-60          &      292.1 &        0.3 &      9.95  &     12.00  &       11.0$\pm$0.1  &   &     4.0$\pm$0.5 & 9.0 & 2.75$\pm$0.34 & \\
4U 1258-61 &      304.1 &       1.2  &      10.9  &      10.6  &       10.0$\pm$0.2 &    &    5.5$\pm$0.7 & 2.4 & 1.82$\pm$0.23 & \\
4U 1254-69 &      303.5 &       -6.4 &      2.16  &      2.91  &       2.6$\pm$0.1  &  2.2$\pm$0.1 &  1.2$\pm$0.1 & 13 & 2.17$\pm$0.55 & 1.83$\pm$0.47 \\
4U 1456-32 &      332.2 &       23.9 &      0.91  &      0.84  &      0.7$\pm$0.1 &  0.9$\pm$0.1 &    0.31$\pm$0.15 & 1.2 & 2.26$\pm$1.14 & 2.90$\pm$1.44 \\
4U 1516-56 &      322.1 &        0.0 &     15.90  &     19.80  &      18.4$\pm$1.2  &    &   12.2$\pm$1.6 & 5.5 & 1.51$\pm$0.22 & \\
4U 1538-52 &      327.4 &        2.2 &      9.14  &      9.70  &      16.2$\pm$0.5  &  20.6$\pm$0.5 &      7.1$\pm$0.7 & 5.5 & 2.28$\pm$0.24 & 2.90$\pm$0.29 \\
4U 1543-47 &      330.9 &        5.4 &      3.51  &      4.00  &       3.8$\pm$0.2  &    &    1.6$\pm$0.2 & 7.5 & 2.38$\pm$0.32 & \\
4U 1543-62 &      321.8 &       -6.3 &      2.43  &      2.96  &       3.2$\pm$0.1  &  3.4$\pm$0.1 &      1.0$\pm$0.3 & 7.0 & 3.20$\pm$0.97 & 3.40$\pm$1.02 \\
4U 1608-52 &      330.9 &       -0.9 &     18.10  &     19.20  &      10.7$\pm$0.2  &  10.8$\pm$2.2 &     6.2$\pm$1.7  & 3.9 & 1.73$\pm$0.47 & 1.74$\pm$0.60 \\
4U 1636-53 &      332.9 &       -4.8 &      2.76  &      3.58  &       2.9$\pm$0.1  &  3.7$\pm$0.1 &      2.5$\pm$0.3 & 6.0 & 1.16$\pm$0.14 & 1.48$\pm$0.18 \\
4U 1658-48 &      338.9 &       -4.3 &      3.74  &      5.26  &       5.0$\pm$0.1  &  8.1$\pm$0.2 &      3.7$\pm$0.4 & & 1.35$\pm$0.15 & 2.19$\pm$0.24 \\
4U 1704-30 &      353.8 &        7.3 &      1.82  &      1.82  &       2.3$\pm$0.2  &    &    0.9$\pm$0.1 & 10.0 & 2.56$\pm$0.36 & \\
4U 1724-307 &     356.3 &        2.3 &      5.70  &      6.57  &       6.6$\pm$0.5  &    &    4.9$\pm$0.6 & 7.4 & 1.35$\pm$0.19 & \\
\end{tabular}
\end{center}
\end{minipage}}
\end{table*}

\begin{table*}
{\scriptsize
\begin{minipage}[b]{1.0\textwidth}
\begin{center}
\begin{tabular}{@{}p{1.9cm}p{0.7cm}p{0.7cm}p{1.4cm}p{1.4cm}p{1.4cm}p{1.4cm}p{0.9cm}p{0.6cm}p{2.0cm}p{1.4cm}@{}}

4U 1728-16 &        8.5 &        9.0 &      1.96  &      2.08  &       2.4$\pm$0.1  &    &    0.8$\pm$0.5 & 4.2 & 3.00$\pm$1.88 & \\
4U 1728-34 &      354.3 &        0.2 &     12.40  &     16.00  &      29.3$\pm$0.5  &    &   10.8$\pm$1.8 & 5.2 & 2.71$\pm$0.45 &\\
4U 1730-335 &     354.8 &       -0.2 &     12.20  &     16.70  &      17.4$\pm$0.3  &     &  10.1$\pm$1.9 & 8.0 & 1.72$\pm$0.19 & \\
4U 1746-37 &      353.5 &       -5.0 &      2.63  &      2.95  &       4.0$\pm$0.3  &     &   1.5$\pm$0.2 & 11.0 & 2.67$\pm$0.41 & \\
4U 1820-30 &        2.8 &       -7.9 &      1.29  &      1.50  &       2.7$\pm$0.1  &  2.6$\pm$0.3  &      0.9$\pm$0.1 & 7.6 & 3.00$\pm$0.35 & 2.89$\pm$0.46 \\
4U 1822-37 &      356.9 &      -11.3 &      1.04  &      1.24  &       1.2$\pm$0.2  &    &    0.5$\pm$0.3 & 2.5 & 2.40$\pm$1.49 & \\
4U 1837+04  &      36.1 &        4.8 &      4.14  &      4.71  &       5.2$\pm$0.2  &    &    1.3$\pm$0.1 & 8.4 & 4.00$\pm$0.34 & \\
4U 1850-08  &      25.4 &       -4.3 &      2.36  &      2.78  &       3.8$\pm$0.1  &  4.8$\pm$0.2 &    1.2$\pm$0.2 & 8.2 & 3.17$\pm$0.53 & 4.00$\pm$0.69 \\
4U 1908+00 &       35.7 &       -4.1 &      2.84  &      3.37  &       4.4$\pm$0.1  &  4.1$\pm$0.1 &      1.2$\pm$0.1 & 5.0 & 3.67$\pm$0.32 & 3.42$\pm$0.30 \\
4U 1915-05 &       31.4 &       -8.5 &      2.31  &      2.60  &       3.9$\pm$0.1  &  5.6$\pm$40.7 &      2.2$\pm$0.9 & 8.9 & 1.77$\pm$0.73 & 2.55$\pm$1.09 \\
4U 1956+35 &       71.3 &        3.1 &      7.21  &      7.21  &       4.4$\pm$0.1  &  5.4$\pm$0.1  &      3.3$\pm$0.1 & 2.14 & 1.33$\pm$0.05 & 1.64$\pm$0.06 \\
4U 1957+11 &       51.3 &       -9.3 &      1.21  &      1.27  &       1.6$\pm$0.1  &  1.1$\pm$0.1  &      0.9$\pm$0.1 & & 1.78$\pm$0.23 & 1.22$\pm$0.18 \\
4U 2129+47 &       91.6 &       -3.0 &      3.30  &      3.82  &       2.4$\pm$0.7  &  3.6$\pm$0.1  &      1.2$\pm$0.2 & 7.0 & 2.00$\pm$0.67 & 3.00$\pm$0.51 \\
4U 2135+57 &         99 &        3.3 &      8.18  &      9.19  &       7.8$\pm$0.3  &    &    4.0$\pm$0.3 & 3.8 & 1.95$\pm$0.16 & \\
4U 2142+38 &       87.3 &      -11.3 &      1.87  &      2.17  &       1.9$\pm$0.1  &  2.9$\pm$0.1 &      1.1$\pm$0.1 & 7.2 & 1.73$\pm$0.18 & 2.64$\pm$0.26 \\
EXO 0331+530 &    146.1 &       -2.2 &      6.91  &      8.54  &       9.0$\pm$1.0  &    &    5.8$\pm$0.2 &  & 1.55$\pm$0.18 & \\
EXO 0748-676 &    280.0 &      -19.8 &      1.01  &      1.14  &       0.8$\pm$0.1  &  0.6$\pm$0.1 &      0.2$\pm$0.1 & 6.8 & 4.00$\pm$2.06 & 3.00$\pm$1.58 \\
EXO 1745-248 &      3.8 &        1.7 &      6.33  &      5.81  &      13.2$\pm$3.5  &    &   7.2$\pm$1.0  & 8.7 & 1.83$\pm$0.55 & \\
EXO 2030+375 &     77.2 &       -1.2 &      8.80  &     10.30  &      20.5$\pm$0.1  &    &   11.7$\pm$0.7 & 7.1 & 1.75$\pm$0.11 & \\
GRO J1719-24 &      0.1 &        7.0 &      2.56  &      2.84  &       4.0$\pm$0.4  &    &   2.8$\pm$0.6  & 1.2 & 1.43$\pm$0.34 & \\
GRO J0422+32 &    165.9 &      -11.9 &      1.55  &      1.66  &       1.6$\pm$0.9  &    &    0.8$\pm$0.1 & & 2.00$\pm$1.15 & \\
GRO J1008-57 &      283 &       -1.8 &     13.50  &     15.10  &      12.7$\pm$0.1  &    &   6.1$\pm$0.2  & 2.0 & 2.08$\pm$0.07 & \\
GRO J1655-40 &      345 &        2.5 &      5.33  &      6.86  &       7.4$\pm$0.3  &  7.4$\pm$0.2 &      3.7$\pm$0.3 & & 2.00$\pm$0.18 & 2.00$\pm$0.17\\
GRO J1944+26 &     63.2 &        1.4 &      9.43  &     10.20  &      13.0$\pm$0.1  &  16.3$\pm$0.3 &     6.5$\pm$0.3  & 9.5 & 2.00$\pm$0.09 & 2.51$\pm$0.12\\
GRO J2058+42 &     83.5 &       -2.7 &       6.22 &        7.1 &      9.5$\pm$2.3   &    &    4.3$\pm$0.4 & 9.0 & 2.21$\pm$0.57 & \\
GRS 1009-45 &     275.9 &        9.3 &      1.05  &      1.16  &       0.4$\pm$0.1  &    &    0.6$\pm$0.2 & 4.5 & 0.67$\pm$0.28 & \\
GS 0834-43 &      262.0 &       -1.5 &     10.10  &     11.80  &      21.2$\pm$2.3  &    &   12.3$\pm$3.7 & 4.0 & 1.72$\pm$0.29 & \\
Gu Mus      &     295.3 &       -7.1 &      1.91  &      2.50  &       2.3$\pm$0.1  &    &    0.9$\pm$0.1 & 5.9 & 2.56$\pm$0.30 & \\
IGR J17544-2619 &   3.2 &       -0.3 &     11.80  &     14.40  &      11.2$\pm$1.2  &    &    6.3$\pm$0.4 & 3.7 & 1.78$\pm$0.22\\
IGR J22534+6243 & 110.0 &        2.9 &      8.96  &      9.98  &       18.1$\pm$1.9 &    &    5.7$\pm$0.4 & & 3.18$\pm$0.40 & \\
KS 1947+300 &      66.1 &        2.1 &      8.90  &     10.40  &       5.1$\pm$0.2  &    &    3.4$\pm$0.4 & 10.0 & 1.50$\pm$0.19 & \\
LS 2883 &         304.2 &       -1.0 &     14.50  &     16.20  &       4.0$\pm$0.3  &    &    3.3$\pm$0.4 & 2.7 & 1.21$\pm$0.17 & \\
LS 5039 &          16.9 &       -1.3 &      8.87  &      8.71  &       6.4$\pm$0.1  &    &    3.7$\pm$0.3 & 2.9 & 1.73$\pm$0.14 & \\
ALS 19596 &        33.0 &        1.7 &     10.10  &     10.10  &      23.0$\pm$1.3  &    &    8.1$\pm$0.9 & & 2.84$\pm$0.35 & \\
MXB 1746-20 &       7.7 &        3.8 &      3.30  &      2.92  &       6.7$\pm$0.3  & 5.7$\pm$0.1   &  3.2$\pm$0.3 & 8.5 & 2.13$\pm$0.22 & 1.78$\pm$0.17 \\
SAX J1748.9-2021 &  7.7 &        3.8 &      3.30  &      2.92  &       7.0$\pm$1.0  &  &    3.2$\pm$0.4 & 8.5 & 2.19$\pm$0.42 & \\
RX J0146.9+6121 & 129.5 &       -0.8 &      7.58  &      8.49  &       4.9$\pm$0.5  &    &    2.9$\pm$0.1 & 2.3 & 1.69$\pm$0.18 & \\
RX J0440.9+4431 & 159.8 &       -1.3 &      5.45  &      6.64  &       3.4$\pm$0.1  &    &    2.2$\pm$0.2 & 3.3 & 1.55$\pm$0.15 & \\
RX J1829.4-2347 &   9.3 &       -6.1 &      1.71  &      1.88  &       2.6$\pm$1.7  &    &    1.2$\pm$0.4 & 6.0 & 2.17$\pm$0.73 & \\
RX J1832-33 &       1.5 &      -11.4 &      0.92  &      1.23  &       0.55$\pm$0.22& 0.58$\pm$0.06 &       0.3$\pm$0.1 & 9.3 & 1.83$\pm$0.95 & 1.93$\pm$0.67\\
Swift J1753.5-0127& 24.9&       12.2 &      1.74  &      1.69  &       2.2$\pm$0.1  & 2.7$\pm$0.1   &       1.3$\pm$0.1 & 6.0 & 1.69$\pm$0.15 & 2.08$\pm$0.18 \\
V404 Cyg   &       73.1 &       -2.1 &      6.66  &      8.10  &       8.1$\pm$0.3  & 8.6$\pm$0.6   &      3.6$\pm$0.5  & 3.5 & 2.25$\pm$0.32 & 2.39$\pm$0.37 \\
XTE J1550-564 &   325.9 &       -1.8 &     10.10  &      8.97  &       6.6$\pm$0.1  & 6.6$\pm$0.6   &       2.9$\pm$0.2 & 5.3 & 2.28$\pm$0.16 & 2.28$\pm$0.16 \\
XTE J1720-318 &   354.6 &        3.1 &      4.77  &      5.25  &      15.1$\pm$0.2  &     &   6.6$\pm$0.6 &  & 2.29$\pm$0.21 & \\
XTE J1808-369 &   355.4 &       -8.1 &      1.13  &      1.34  &       1.4$\pm$0.1  & 1.4$\pm$0.1   &        0.8$\pm$0.3 & 3.0 & 1.75$\pm$0.67 & 1.75$\pm$0.67\\
XTE J1819-254 &        6.8 &       -4.8 &       1.81 &       2.34 &                    & 2.4$\pm$0.1 &      0.8$\pm$0.1 & 9.5 & & 3.00$\pm$0.63 \\
XTE J1859+226 &       54.0 &        8.6 &      2.16  &      2.21  &       3.6$\pm$0.4  &   &    1.8$\pm$0.4 & 7.6 & 2.00$\pm$0.50 & \\
XTE J2103+457 &       87.1 &       -0.7 &      6.64  &      7.78  &       7.6$\pm$0.1  &   &    4.2$\pm$0.3 & 6.5 & 1.81$\pm$0.13 & \\
XTE J2123-058 &       46.5 &      -36.2 &      0.54  &      0.57  &       0.8$\pm$2.3  &   &    0.37$\pm$0.15 & 8.5 & 2.16$\pm$6.28 & \\
\hline
\end{tabular}
\end{center}
\begin{tabular}[t]{l}
$^a$ Gl: Galactic longitude; Gb: Galactic latitude.\\
$^b$ KBH: HI column density at the source (column one) direction based on Kalberla, Burton, Hartmann et al.\ (2005).\\
$^c$ DL: HI column density at the source (column one) direction based on Dickey \& Lockman (1990).\\
$^d$ $\NH({\rm AG89})$: $\NH$ derived with the solar abundances of AG89.\\
$^e$ $\NH({\rm W00})$: $\NH$ derived with subsolar abundances
                                       (see Wilms et al.\ 2000 and references therein).\\
$^f$ D: Distance.
\end{tabular}
\end{minipage}
}
\end{table*}
%%%% Table 4 %%%%

%%%% Figure 2 %%%%
\begin{figure*}
\centerline{\includegraphics[width=0.49\textwidth, angle=0]{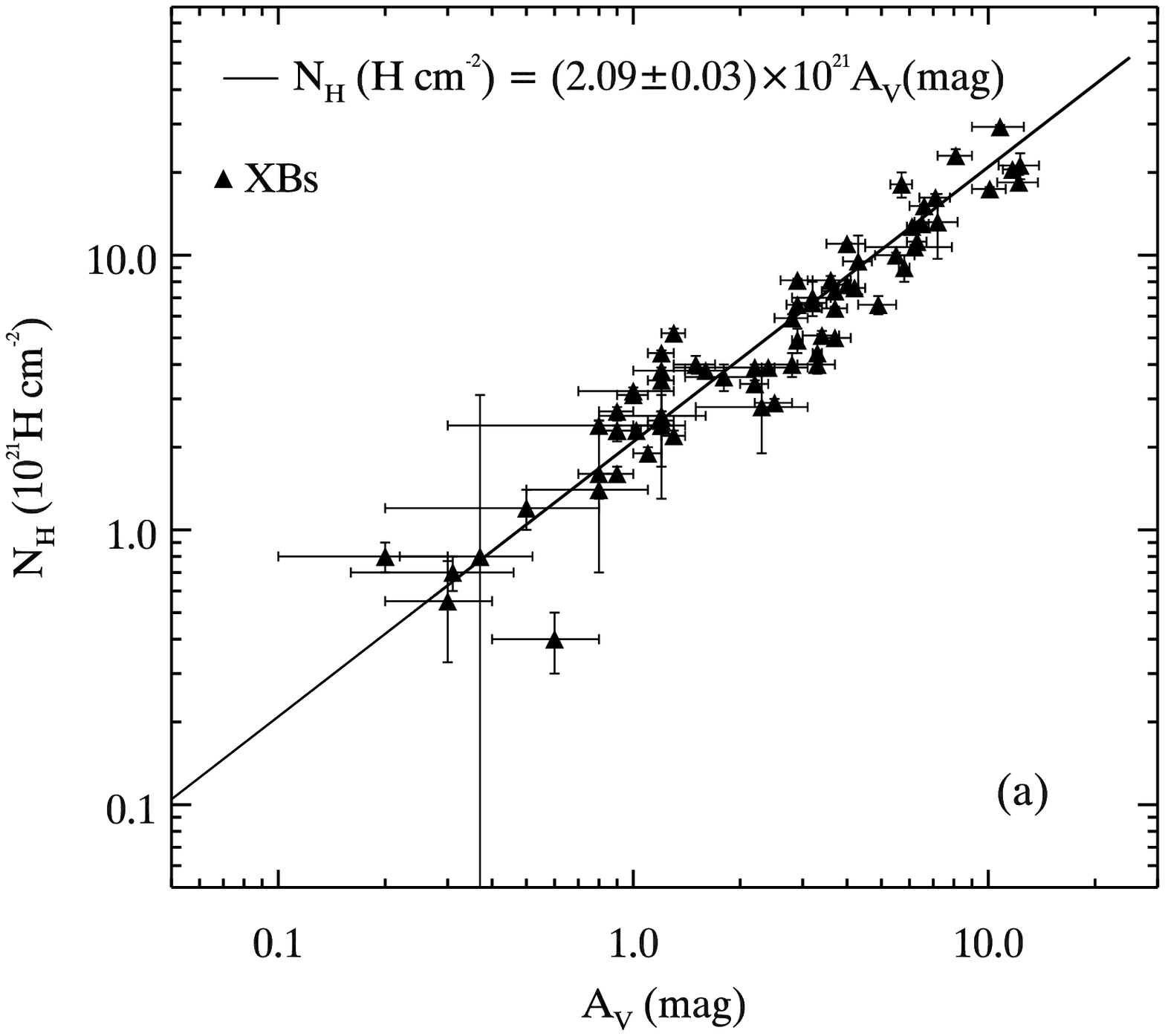}
\includegraphics[width=0.49\textwidth, angle=0]{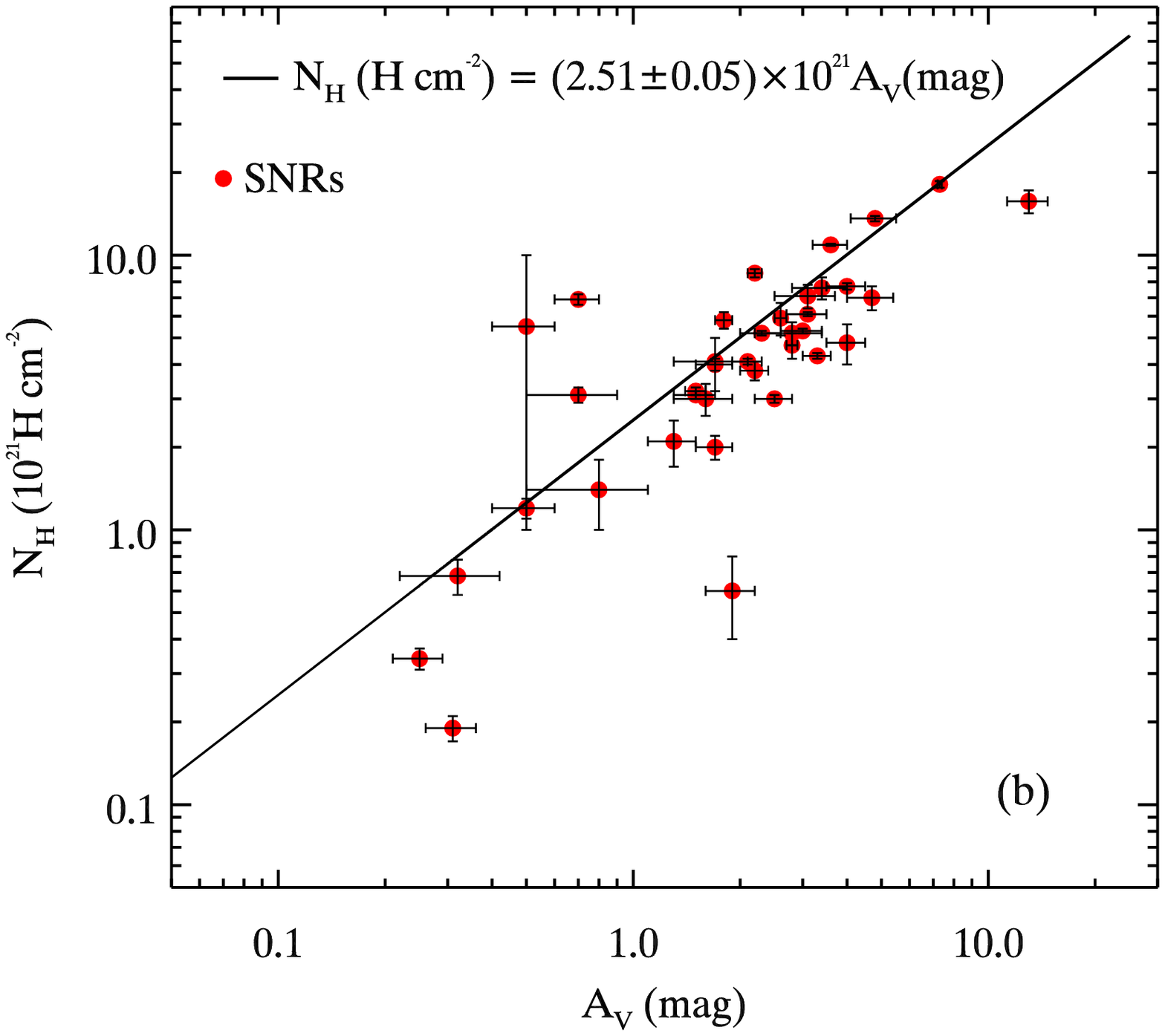}}
\caption{Left (a): $\NH$--$\AV$ relation derived from
               the XBs of the AG89 sample.
               Right (b): Same as (a) but for SNRs.
               }
\label{fig2}
\end{figure*}
%%%% Figure 2 %%%%

%%%% Figure 3 %%%%
\begin{figure*}
\centerline{\includegraphics[width=0.49\textwidth, angle=0]{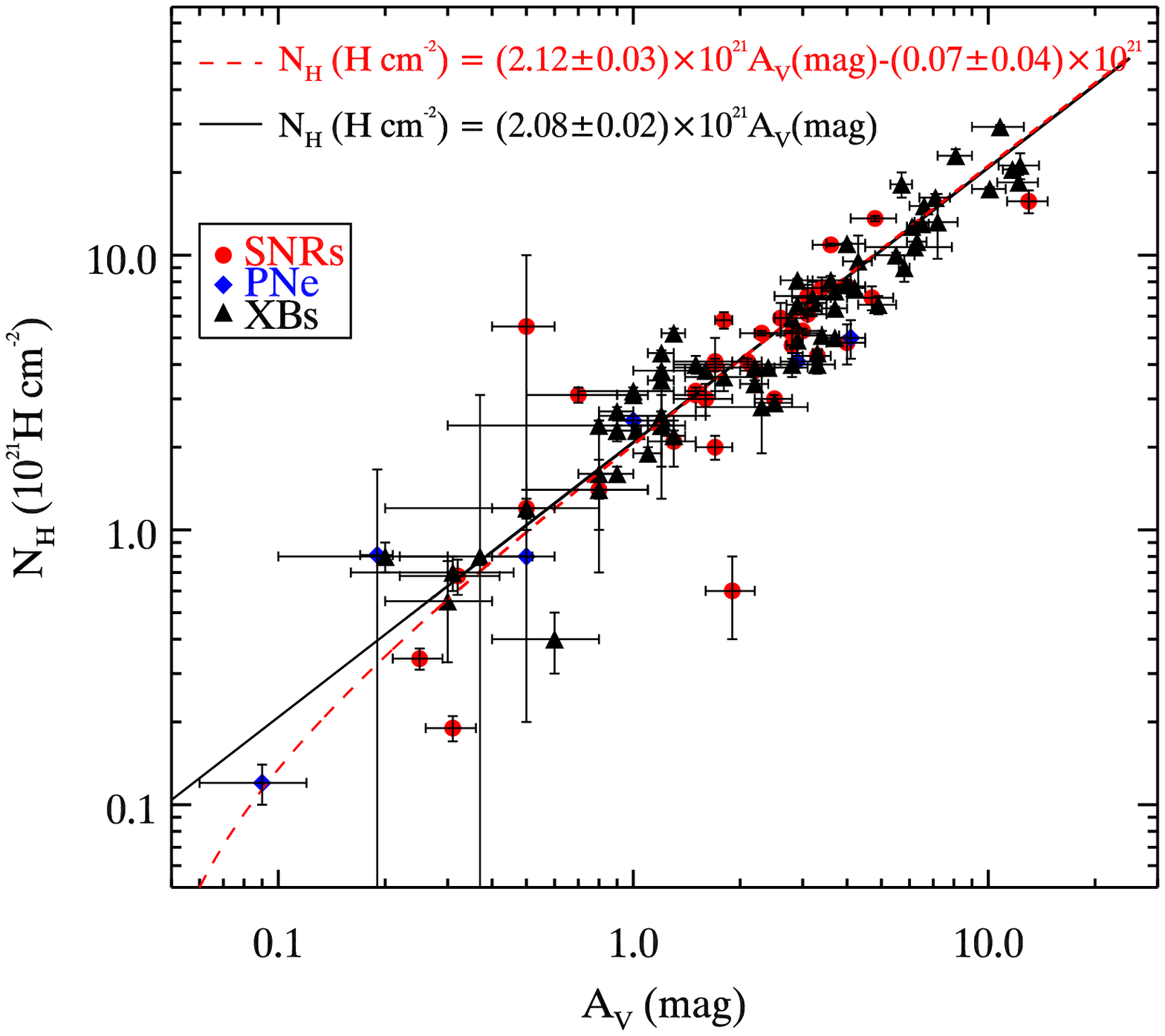}
\includegraphics[width=0.49\textwidth, angle=0]{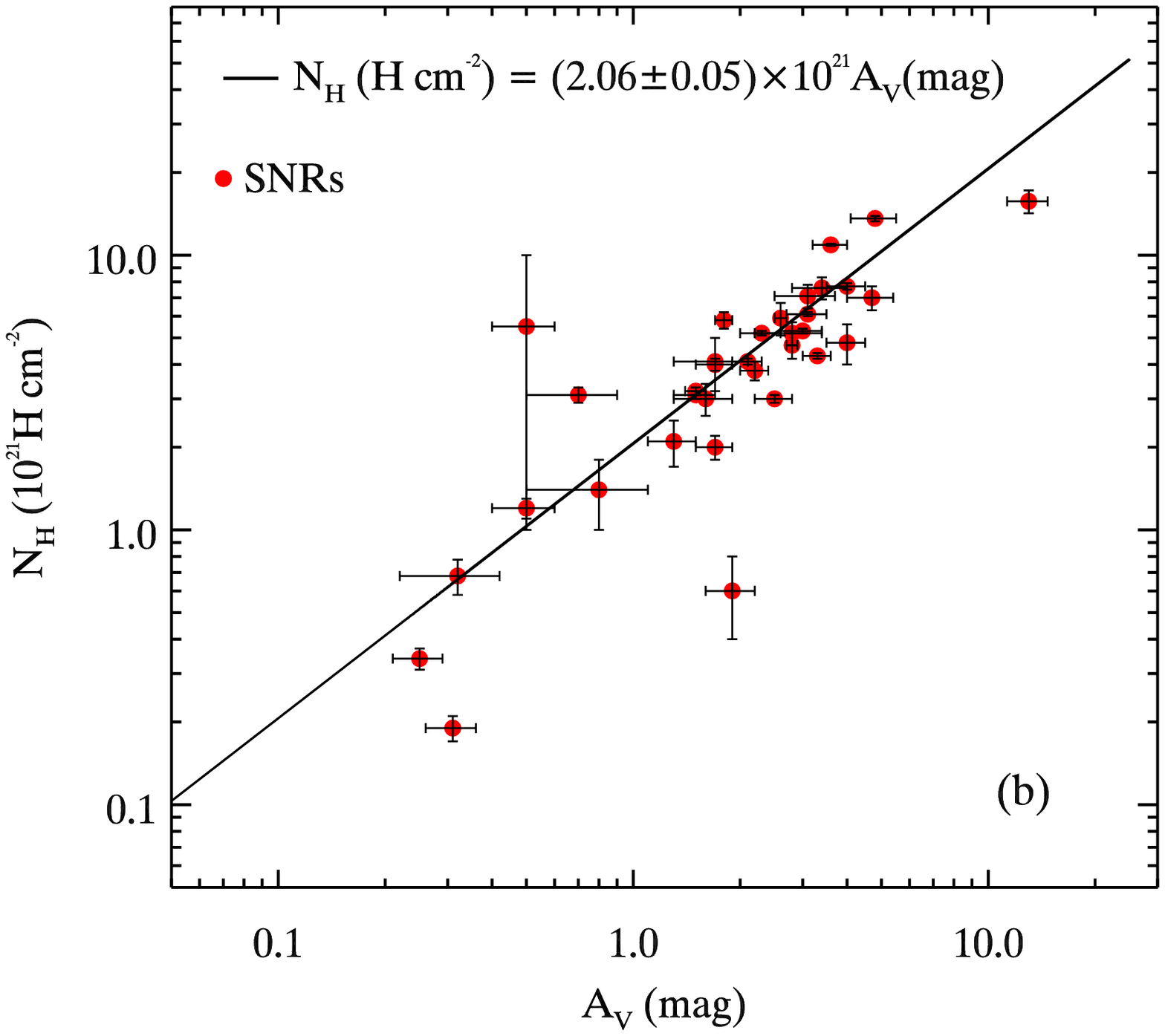}}
\caption{Left (a): $\NH$--$\AV$ relation derived for the whole AG89 sample
               but with the three outlying SNRs excluded.
               Right (b): Same as Figure~2b but with
               the three outlying SNRs excluded.
               }
\label{fig3}
\end{figure*}
%%%% Figure 3 %%%%

Bohlin et al.\ (1978) derived
the hydrogen-to-extinction ratio from
a sample of 100 stars with $E(B-V)$ up to $\simali0.5\magni$
based on the ultraviolet (UV) absorption spectra
of HI and H$_2$ observed by the {\it Copernicus} satellite.
With $\RV\approx3.1$ for the Milky Way (MW)
diffuse interstellar medium (ISM; Whittet 2003),
the Bohlin et al.\ (1978) hydrogen-to-extinction ratio becomes
$\NH/\AV\approx1.87\times10^{21}\rmH\cm^{-2}\magni^{-1}$.
Similarly, as summarized in Table~1,
Jenkins \& Savage (1974), Whittet (1981),
Diplas \& Savage (1994), and Rachford et al.\ (2009)
derived the hydrogen-to-extinction ratio
from the UV absorption spectra of HI and H$_2$
measured by the {\it Orbiting Astronomical Observatory} 2
(OAO-2; Jenkins \& Savage 1974),
the {\it Copernicus} satellite (Whittet 1981),
the {\it International Ultraviolet Explorer}
(IUE; Diplas \& Savage 1994),
and the {\it Far Ultraviolet Space Explorer}
(FUSE; Rachford et al.\ 2009).
The $\NH/\AV$ ratios have also been derived
by Sturch (1969), Knapp \& Kerr (1974),
Heiles (1976), Liszt (2014),
and Chen et al.\ (2015) based on
the column densities of hydrogen
measured by CO and
the 21\,cm emission of HI.

While the aforementioned studies directly measured
the column densities of HI and H$_2$
(see Table~2),
an alternative way of determining $\NH$
is to measure the X-ray absorption spectra
of heavy elements such as O, Mg, Si and Fe.
One derives the column densities of
these heavy elements from the measured X-ray spectra
and then converts to $\NH$
by assuming an interstellar reference abundance standard
which specifies the relative
abundances of these heavy elements
(i.e., O/H, Mg/H, Si/H, and Fe/H).
This method take ionized, neutral and molecular hydrogen into consideration
and has been applied to measure $\NH$ toward
supernova remnants (SNRs), X-ray binaries (XBs),
star-forming (SF) regions, and extended sources
(see Table~3).

However, the $\NH/\AV$ ratios derived from
the X-ray absorption data involves the unknown
interstellar abundance X/H
(where X = O, Mg, Si, Fe ...),
often expressed in units of ppm
(i.e., parts per million).
One often assumes the interstellar elemental
abundance to be solar. However, the solar abundances
of the key heavy elements such C, O, N, S, Mg, Si,
and Fe reported in the literature have undergone
substantial changes over the past four decades
(see Li 2005, Asplund et al.\ 2009). Also, there
are arguments that the interstellar abundances
could be better represented by those of
B stars (because of their young ages)
which are just $\simali$60--70\% of the widely
adopted solar values (``subsolar'';
see Snow \& Witt 1995, 1996, Sofia \& Meyer 2001).
As shown in Table~3, from the same set of X-ray data
one would derive a much higher (by $>30\%$) $\NH/\AV$ ratio
if one adopts the reference abundance
of Wilms et al.\ (2000; hereafter W00)
for O and Fe
instead of that of Anders \& Grevesse
(1989; hereafter AG89).
Also, these previous studies suffer from small sample sizes
and/or limited space distributions of the sample sources.

To reduce the above-mentioned limitation
and to obtain a more reliable $\NH/\AV$ ratio,
in this work we carefully compile $\NH$
and $\AV$ from the literature
by separating those with $\NH$
derived from solar abundances (AG89)
from those with $\NH$ derived from
subsolar abundances (W00).
We finally arrive at a sample of
35 SNRs, 6 planetary nebulae (PNe),
70 XBs for which $\NH$ was derived
with solar abundances (AG89)
and 19 SNRs, 29 XBs for which
$\NH$ was derived with
subsolar abundances (W00).
%
%in order to reduce the limitation and
%obtain a more reliable $\NH/\AV$ ratio.
We present our sample in \S2
and derive a renewed mean
$\langle\NH/\AV\rangle$ ratio in \S3.
Also in \S3 we estimate the gas-to-dust mass ratio
and explore the spatial variation of
$\langle\NH/\AV\rangle$
and the gas distribution in the Galaxy.
The major results are summarized in \S4.

%%%
\section{Compilation of $\AV$ and $\NH$}
%%%
\subsection{$\AV$}
%%%
If a group of lines are from the transitions
with the same upper level, their intensity ratios
would only depend on the transition probabilities.
Thus they can be employed to calculate the extinction
toward extended objects, e.g., SNRs and PNe.
The frequently used line ratios include
H$\alpha$(6563$\Angstrom$)/H$\beta$(4861$\Angstrom$),
known as the Balmer decrement
(e.g., SNR G67.7+1.8, Mavromatakis et al.\ 2001),
[SII]($\sim$10320$\Angstrom$)/[SII]($\sim$4068$\Angstrom$)
(e.g., SNR G111.7-2.1, Hurford \& Fesen 1996)
and [FII]($\sim$16435$\Angstrom$)/[FII]($\sim$12567$\Angstrom$)
(e.g., SNR G332.4-0.4, Oliva et al.\ 1989).
However, caution needs to  be taken
when applying the Balmer decrement method
to very young SNRs, because their intrinsic
H$\alpha$/H$\beta$ ratios may be different
from the canonical ratios used for nebulae
under the ``Case B'' condition,
in which the nebula is always optical thin
except for Lyman-line photons
(e.g., Tycho, Ghavamian et al.\ 2000).
Other methods to estimate the extinction of SNRs
include the use of existing extinction maps
(e.g., SNR G119.5+10.2, Mavromatakis et al.\ 2000),
and the association of the targets
with extinction-known objects
(e.g., SNR G260.4-3.4, Gorenstein 1975).

For XBs, four methods are usually adopted
to estimate their extinction:
(1) the relation between the colour excess
    or reddening $E(B-V)$ and the widths of
    the diffuse interstellar bands
    or Na doublet lines
    (e.g., 4U0614+091, Nelemans et al.\ 2004);
(2)  the colour-magnitude diagram
      for those in star clusters
      (e.g., 4U 1820-30 in NGC 6624, Valenti et al.\ 2004);
(3) a comparison of the observed colour with the
    intrinsic colour derived from the stellar spectral type
    (e.g., 4U1908+005, Thorstensen et al.\ 1978); and
(4) fitting the observed, reddened UV spectrum
     of a source (e.g., 2S 0620-003)
     to determine the excess extinction
     at 2175$\Angstrom$ ($\Delta A_{\rm bump}$) and
     assuming a constant $\Delta A_{\rm bump}/\AV$ ratio.

For each source in our sample
we estimate its $\AV$ or $E(B-V)$\footnote{%
   From  $E(B-V)$ to $\AV$ we take $\AV = 3.1\times E(B-V)$.
   }
following these rules:
(1) if a source has more than one measure
     of $\AV$ or $E(B-V)$,
    an error-weighted value will be adopted;
(2) if a source has more than two measures
     of $\AV$ or $E(B-V)$ and one measure
     differs from the others
     by $\simgt$30\% or 3$\sigma$,
     this measure will be rejected;
(3) if there is no error estimation
     for the $\AV$ or $E(B-V)$ measure,
     we will assign an uncertainty of
     $\simali$13\% which is obtained
     by averaging over those sources
     for which the uncertainties on
     $\AV$ or $E(B-V)$ are known
     (see Column 13 in Table~7).
%

%%%
\subsection{$\NH$}
%%%
The X-ray absorption factor is usually expressed as
%
%%%% Equation 1 %%%%
\begin{equation}
M(E) = \exp\left\{{\NH}{\sigma_{\rm eff}}(E)\right\},
\end{equation}
%%%%% Equation 1 %%%%
%
where $\sigma_{\rm eff}(E)$ represents
the effective absorption cross section taking into
account the composition of interstellar medium.
Apparently, the interstellar X-ray absorption
is affected by both $\NH$ and $\sigma_{\rm eff}(E)$.
To be self-consistent, it is necessary
to seek $\NH$ from the literature
using similar heavy element abundances
(e.g., O, Ne, Fe, Mg, Si).
As mentioned earlier, different abundance selection
(e.g., AG89 vs W00)
may lead to a difference of $>30\%$ for the derived
$\NH$ (Foight et al.\ 2016).
In this study, we will consider
the hydrogen column densities
derived either from solar abundances (AG89)
or subsolar abundances (W00).
If there is no mention in the literature
about the abundance setting
for the software {\it XSPEC},
we will assume that the AG89
abundance was used because it is the default setting
at least back to {\it XSPEC} version 8.

We take the following approach for adopting $\NH$:
(1) if a source has more than one measure of $\NH$,
     an error-weighted value will be adopted;
(2) if there are more than one measure of $\NH$
     and these measures {\it all} differ from each other
     by $\simgt$30\% or 3$\sigma$,
     the one derived from the data
     with a higher quality at energy $E\simlt2\keV$
     will be selected because this energy range is
     more sensitive to X-ray absorption;
    this means that those derived
     from {\it Chandra}, {\it XMM-Newton}
     and {\it Suzaku} have a higher priority than
     those from {\it ROSAT}, {\it RXTE}
     and other telescopes;
(3) if there are two or three measures
      of $\NH$ for the same source
      and they are all derived from {\it Chandra},
      {\it XMM-Newton} or {\it Suzaku},
      we will take the error-weighted value
      without considering their differences;
(4) if there is no error estimation for a $\NH$ measure,
     we will adopt an uncertainty of
     $\simali$9\%  for AG89 or
     $\simali$8\% for W00
     which are obtained
     by averaging over those sources
     for which the uncertainties on $\NH$
     are known
     (see Columns 5, 9 in Table~7);
(5) for SNRs which are extended sources
     of which $\AV$ and $\NH$
     are known for a range of regions,
     we make sure that the adopted $\AV$
     and $\NH$ are from the same region, if possible.
Note that for XBs, some of them are dipper sources
and have intrinsic local absorptions.
We try to avoid these local absorptions
by taking the following procedure:
(1) for low mass X-ray binaries (LMXBs),
we examine the catalog of Liu et al.\ (2006)
to see whether they are dipper sources or not;
if yes, we will exclude them from our sample
except those for which $\NH$ was derived from
the spectral fitting and no dipper shows up
in the spectra or the interstellar $\NH$ and
local $\NH$ are considered separately
during the fitting
(e.g., using the partial covering model);
(2) for high mass X-ray binaries (HMXBs),
we examine them one by one through the literature
to see whether the interstellar $\NH$ and local $\NH$
are considered separately and whether the interstellar
$\NH$ varies during different phases and states
to avoid the influence of local absorptions.
With these selection procedures,
seven XBs (4U 0115+63, HD 259440,
IGR J11215-5952, IGR J18450-0435,
IGR J2000.6+3210, IGR J21343+4738,
and LS 992) are exclused due to the possible
presence of local absorptions.
For three XBs (GRO J2058+42, GS 0834-43,
and ALS 19596), we find no evidence for
or against the presence of considerable
amounts of local absorption.
As their $\NH/\AV$ ratios are not much
larger than those previously published,
we decide to keep them in our sample.
We also note that, although SNRs are treated
as having little or no intrinsic absorption,
some of them may actually have
considerable local dust emission
(e.g.,  Cas A [Dunne et al.\ 2009]).

In Table 4, we present the finally adopted data.
For each source we also list the $\NH$ value
determined from the HI 21\,cm observation
toward their lines of sight.
One can find all the compiled $\NH$ and $\AV$
for each sources in Appendix.
%

%%%% Figure 4 %%%%
\begin{figure*}
\centerline{\includegraphics[width=0.7\textwidth, angle=0]{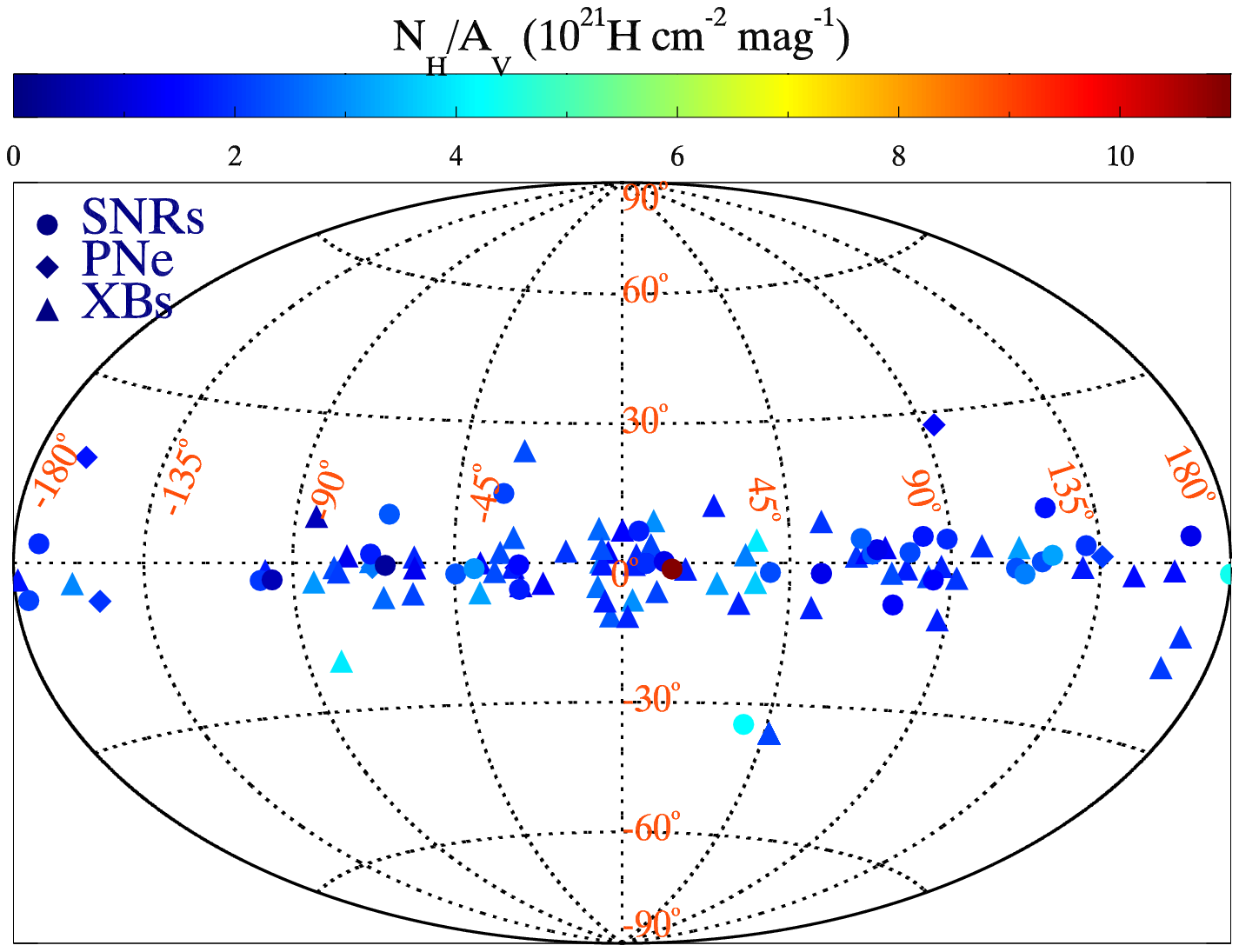}}
\centerline{\includegraphics[width=0.7\textwidth, angle=0]{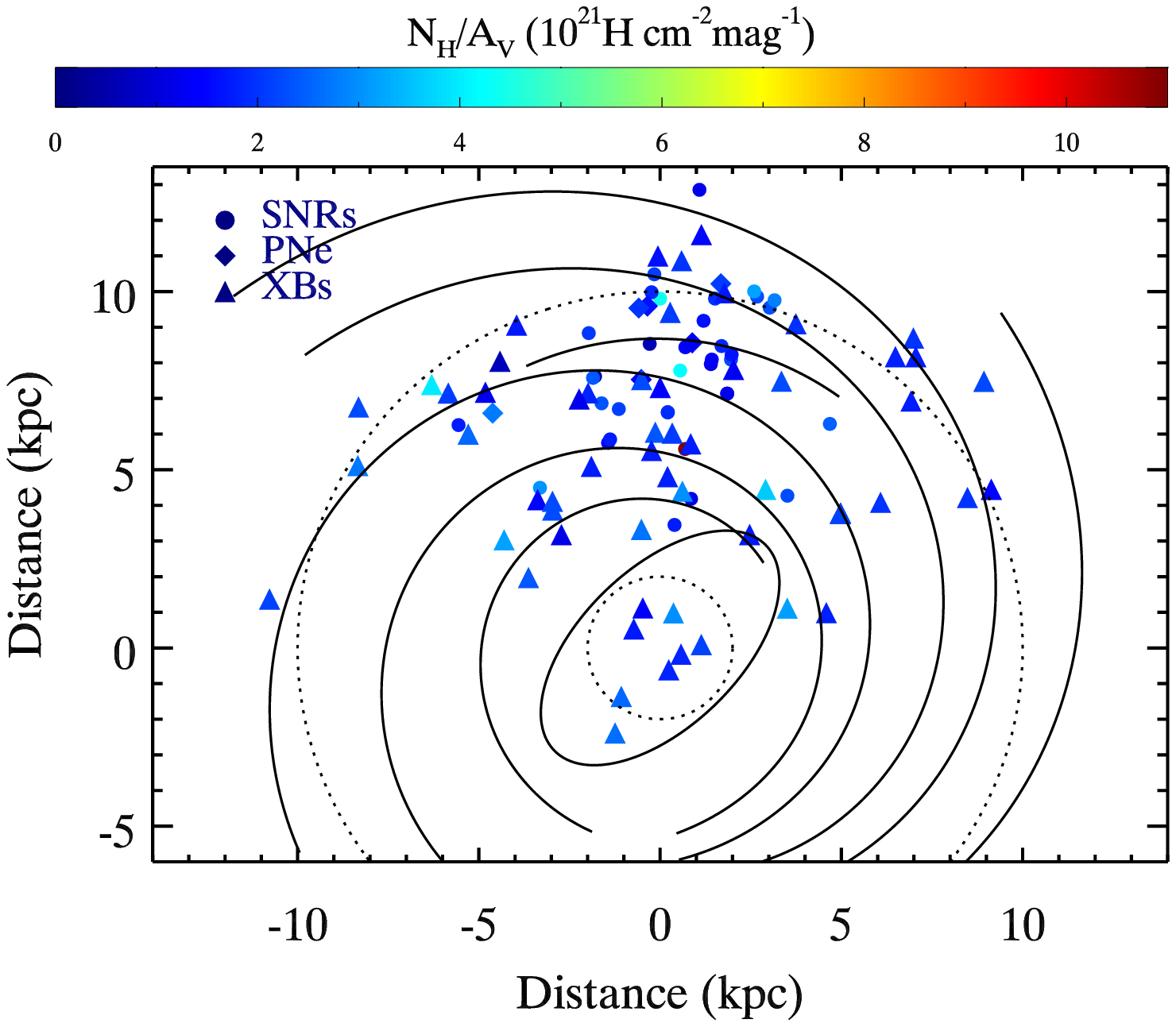}}
\caption{The distribution of $\NH$/$\AV$ of the AG89 sample projected on
         the Galactic longitude-latitude plane (upper panel)
         and the Galactic disk (lower panel).
         In the lower panel, the Galactic centre is marked by a plus sign.
         The two dot circles indicate 2\,kpc and 10\,kpc
         away from the Galactic centre. The solid lines present the spiral arms
         from Hou \& Han (2014) and Green et al.\ (2011).}
\label{fig4}
\end{figure*}
%%%% Figure 4 %%%%

%%%% Figure 5 %%%%
\begin{figure*}
\centerline{\includegraphics[width=0.445\textwidth, angle=0]{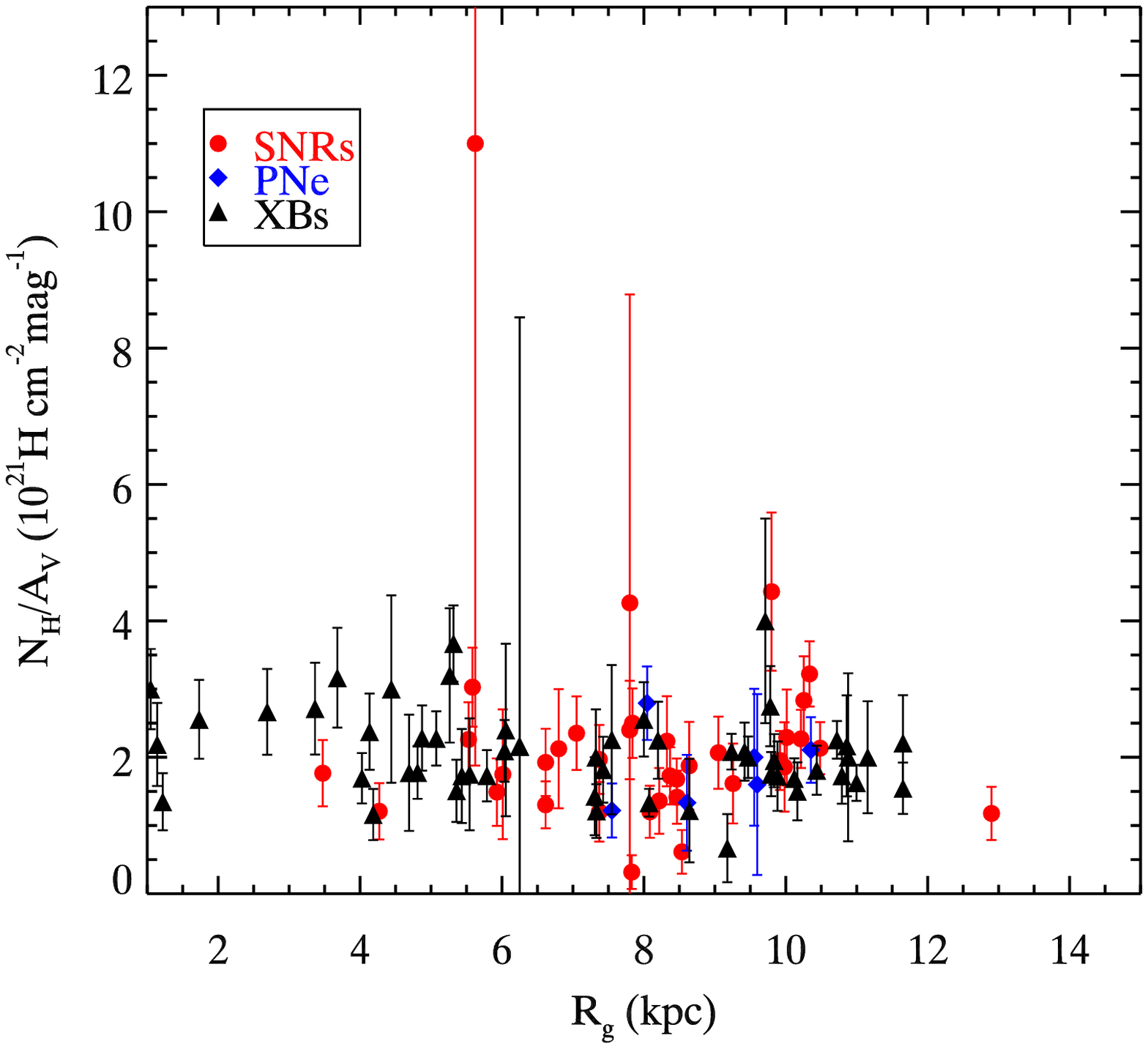}
\includegraphics[width=0.46\textwidth, angle=0]{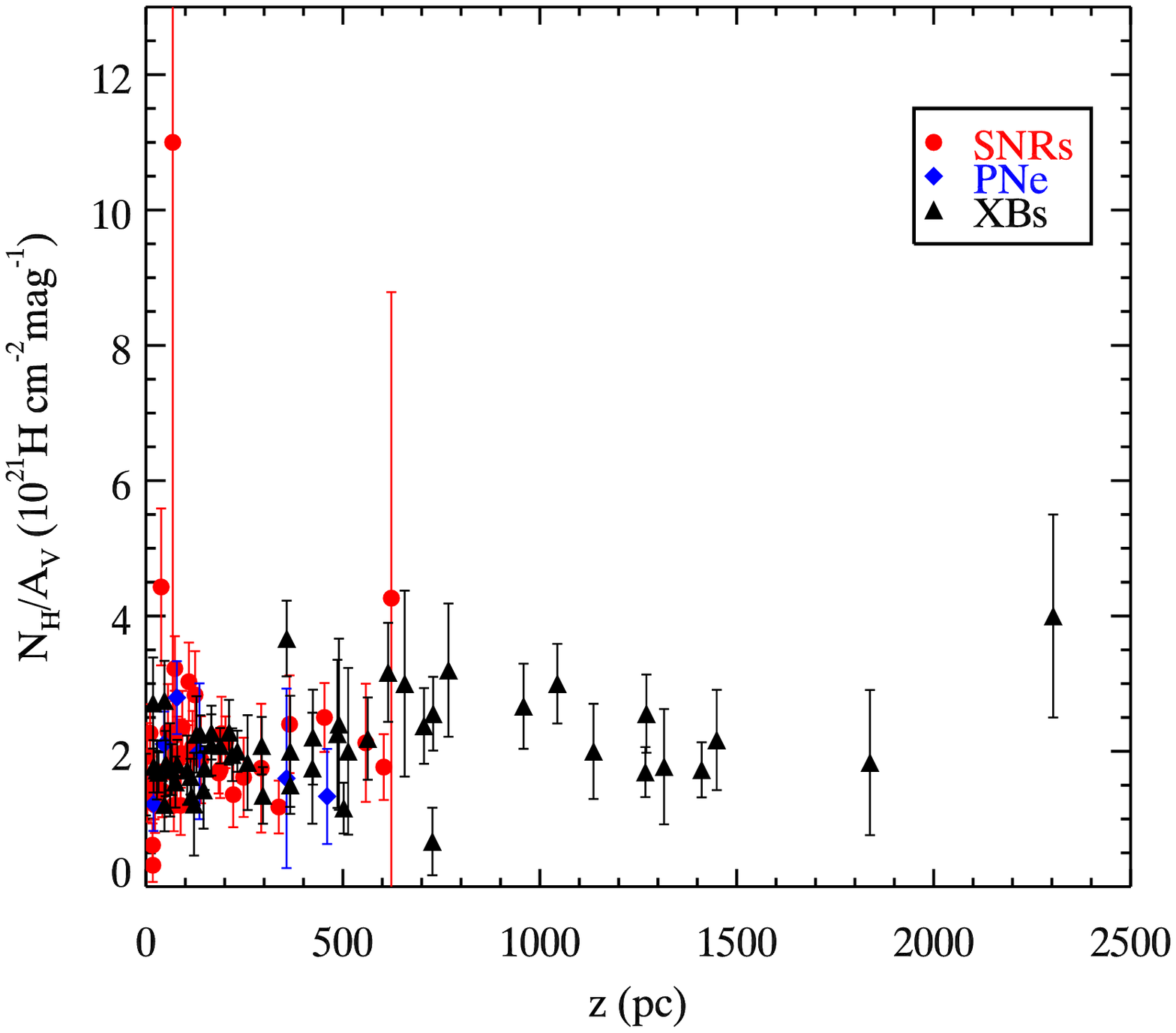}}
\centerline{\includegraphics[width=0.45\textwidth, angle=0]{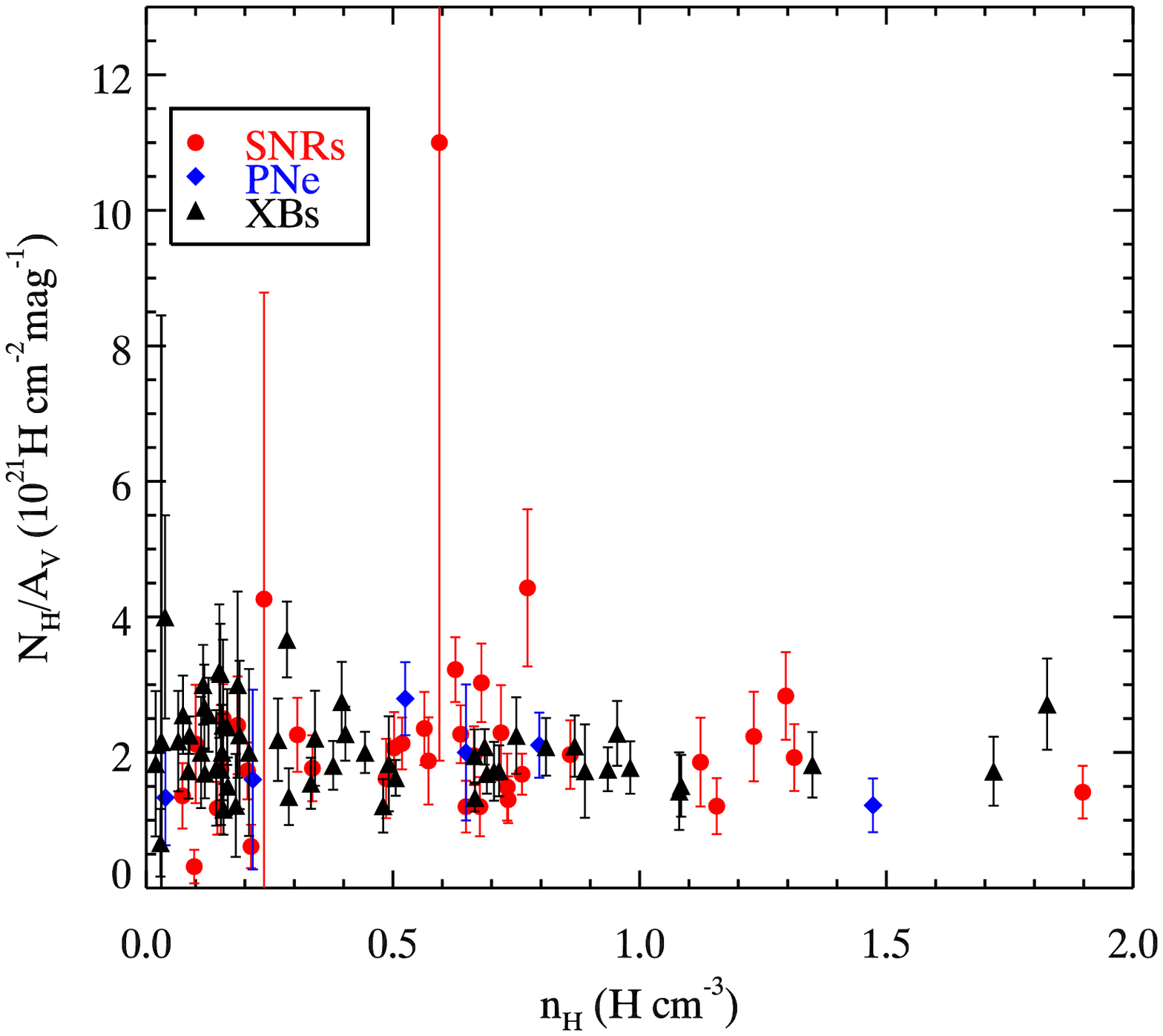}
\includegraphics[width=0.45\textwidth, angle=0]{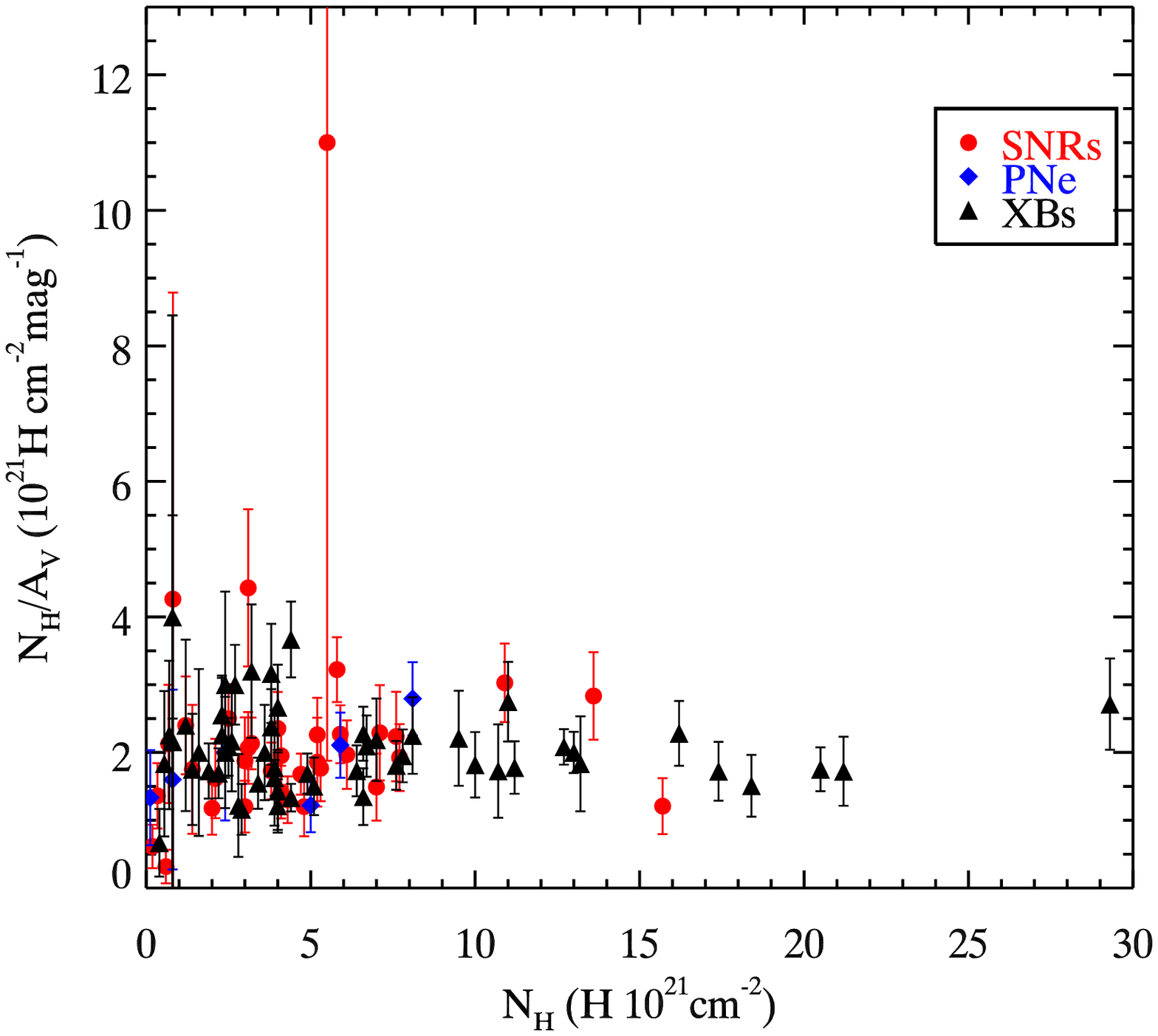}}
\caption{Variations of $\NH/\AV$ with
              $R_g$, $z$, $\nH$ and $\NH$.}
\label{fig5}
\end{figure*}
%%%% Figure 5 %%%%

%%%% Figure 6 %%%%
\begin{figure*}
\centerline{\includegraphics[width=0.45\textwidth, angle=0]{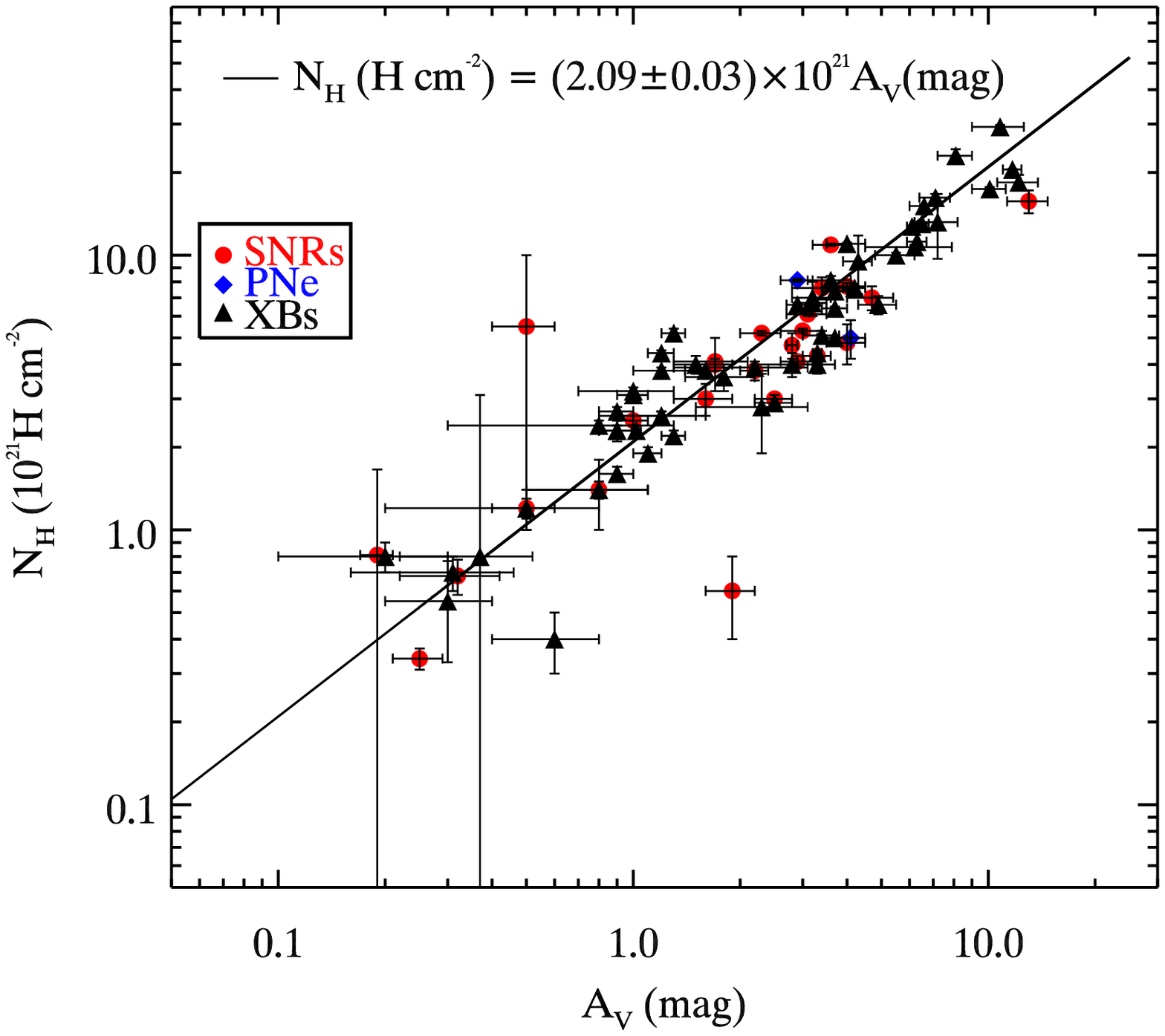}
\includegraphics[width=0.45\textwidth, angle=0]{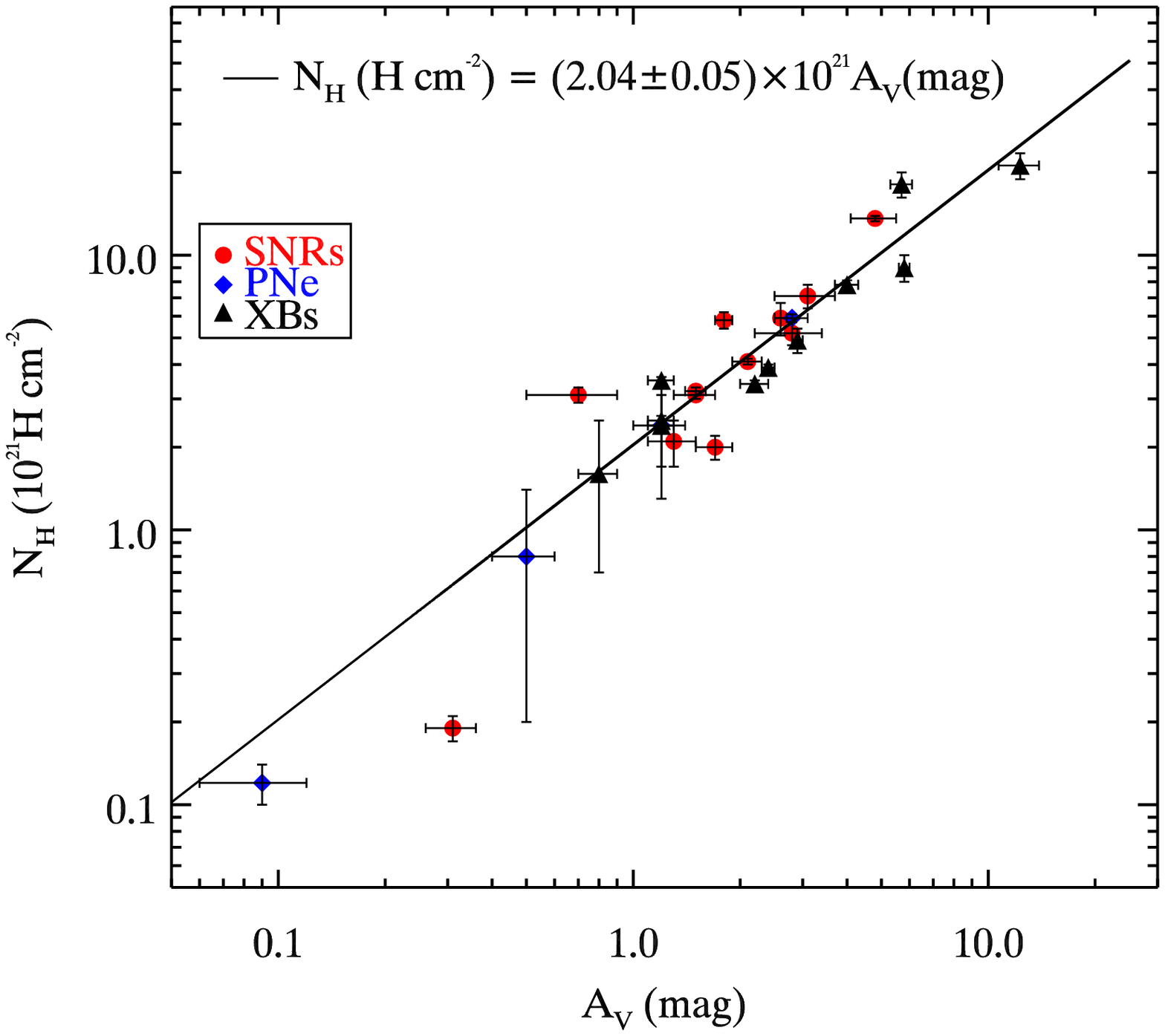}}
\caption{$\NH$--$\AV$ relations
         for the Galactic 1st/4th quadrants (left)
         and for the Galactic 2nd/3rd quadrants (right).
         }
\label{fig6}
\end{figure*}
%%%% Figure 6 %%%%

%%%
\section{Results and Discussion}
%%%
As the AG89 sample
(i.e., those sources for which $\NH$
was derived from solar abundances)
is nearly three times larger than
the W00 sample
(i.e., those sources for which $\NH$
was derived from subsolar abundances),
the former sample allows us not only to
determine $\NH/\AV$
but also to investigate the variation of $\NH/\AV$
and the distribution of interstellar gas in our Galaxy.
Therefore, in this section, we will first focus on
the AG89 sample  in \S3.1--\S3.4.
We will consider the W00 sample in \S3.5.
The possible caveats of our study will be discussed
in \S3.6.

\subsection{The mean $\NH$--${\AV}$ relation}
%%%
Figure~1 plots $\NH$ against $\AV$
for all calibration points with abundance of AG89.
We fit a linear relationship to the data using the
MPFITXY routine (Williams, Bureau \& Cappellari 2010)
of the MPFIT package (Markwardt 2009):\\
%
%%%% Equation 2 %%%%
\begin{equation}
\NH (\rmHAL\cm^{-2}) = \left(2.19\pm0.02\right)
\times10^{21}\,\AV (\magniAL),
\end{equation}
%%%% Equation 2 %%%%
%
where the errors represent a 1$\sigma$ statistical uncertainty.
The black solid lines in Figure 1 shows the best-fit result.
To assess the influence of possible intrinsic absorption in our sample,
we add a zero point offset to equation (2) and we then obtain a slightly
steeper relation as shown as the red dashed line in Figure~1.\\
%
%%%% Equation 3 %%%%
\begin{equation}
\begin{array}{l}
\NH (\rmHAL\cm^{ - 2}) = \left(2.25\pm0.03\right)
\times 10^{21}\,\AV (\magniAL) \\
\qquad \qquad \qquad
- \left(0.11\pm0.04\right)
\times 10^{21} (\rmHAL\cm^{ - 2}) ~~.\\
\end{array}
\end{equation}
%%%% Equation 3 %%%%
%
Since the zero point offset in equation (4)
is much smaller than the column density of most of
our calibration points, it indicates that our sample is not
affected significantly by the possible intrinsic absorption.
%
%%%% Table 5 %%%%
\begin{table}
%\scriptsize
\centering
\caption{A summary of the final $\NH$-$\AV$ fitting.}
\begin{tabular}[t]{ccc}
\hline
\hline
Sample & Coefficient
            & Zero Offset\\
       &($10^{21}\rmH\cm^{-2}\magni^{-1}$)
       & ($10^{21}\rmH\cm^{-2}$)\\
\hline
       & \multicolumn{2}{c}{AG89} \\
Whole      & 2.08$\pm$0.02 & 0 (fixed)\\
Whole      & 2.12$\pm$0.03 & $-0.11\pm0.04$\\
SNRs       & 2.06$\pm$0.05 & 0 (fixed)\\
XBs        & 2.09$\pm$0.03 & 0 (fixed)\\
       & \multicolumn{2}{c}{W00} \\
Whole      & 2.47$\pm$0.04 & 0 (fixed)\\
Whole      & 2.59$\pm$0.07 & $-0.23\pm0.10$\\
SNRs       & 2.51$\pm$0.10 & 0 (fixed)\\
XBs        & 2.46$\pm$0.05 & 0 (fixed)\\
%1st/4th quadrant & 2.07$\pm$0.05 & 0 (fixed)\\
%2nd/3rd quadrant & 2.28$\pm$0.12 & 0 (fixed)\\
\hline
\end{tabular}
\label{tab:5}
\end{table}
%%%% Table 5 %%%%

We also explore whether different kinds of calibrator
affect the derived $\NH$-$\AV$ relation.
To this end, we fit the data of SNRs and XBs separately
and obtain
%%%% Equation 4 %%%%
\begin{equation}
{\NH} ({\rmHAL}{\cm}^{-2}) = \left(2.51{\pm}0.05\right)
{\times}10^{21}\,\AV ({\magniAL})
\end{equation}
%%%% Equation 4 %%%%
for SNRs, and
%%%% Equation 5 %%%%
\begin{equation}
{\NH} ({\rmHAL}{\cm}^{-2}) = \left(2.09{\pm}0.03\right)
{\times}10^{21}\,{\AV} ({\magniAL})
\end{equation}
%%%% Equation 5 %%%%
for XBs.
Surprisingly, they are inconsistent
with each other and the mean
$\langle\NH/\AV\rangle$ ratio
derived from SNRs is larger by $\simali$20\%
than that from XBs.
To find what causes this difference,
we show the $\NH$--$\AV$ relations
in Figure~2 for SNRs and XBs, separately.
Apparently, those three SNRs
(G54.1+0.3, G85.9-0.6, and G132.7+3.1)
which lie well above the best-fit line
(see Figure~2, right panel) lead to
an overestimation of $\NH/\AV$.
Indeed, for G54.1+0.3,
Kim et al.\ (2013) estimated
$\Av\sim7.3\pm0.1\magni$
based on the extinction toward
several stars associated with the SNR,
while the extinction toward
the majority of the stars
is in the range from $\simali$6.9
to $\simali$7.5$\magni$
and for one star $\AV$
is as high as $\simali$7.8 $\magni$
(see their Table~3).
Also, Koo et al.\ (2008) derived an extinction
of 8.0$\pm$0.7$\magni$.
Therefore, an uncertainty of 0.1$\magni$
may have been underestimated.
For G85.9-0.6, G{\"o}k et al.\ (2009)
obtained $\Av$=0.47$\pm$0.05$\magni$,
1.18$\pm$0.12$\magni$,
and 1.40$\pm$0.14$\magni$
at three positions which are not fully overlapped
with the X-ray bright region.
Therefore, the error-weighted $\Av$
and the associated uncertainty that we used
may be not a good estimation
for the X-ray bright region.
For G132.7+3.1, the optical spectroscopy
was taken at the western filaments
(Fesen et al.\ 1995).
However, only its central and northern regions
are bright at X-ray.
While the size of G132.7+3.1 is
about 80$^{\prime}$,
the space separation between
the optical and the X-ray observations
could be larger than 0.5$^{\rm o}$.
Taking these concerns into consideration,
we decide to exclude these three SNRs.
We then obtain
%%%% Equation 6 %%%%
\begin{equation}
{\NH} ({\rmHAL}{\cm}^{-2}) = \left(2.06{\pm}0.05\right)
{\times}10^{21}\,\AV ({\magniAL})
\end{equation}
%%%% Equation 6 %%%%
for SNRs, and
%%%% Equation 7 %%%%
\begin{equation}
{\NH} ({\rmHAL}{\cm}^{-2}) = \left(2.08{\pm}0.02\right)
{\times}10^{21}\,{\AV} ({\magniAL})
\end{equation}
%%%% Equation 7 %%%%
for the whole AG89 sample.
We present the best-fit lines in Figure~3.
Table~5 summarizes the above results.
Our results on the $\NH/\AV$ ratio with AG89 abundance
are compatible with those previously obtained from X-ray data
with the same abundance.
%

%%%%
\subsection{A constant $\NH/\AV$ ratio across the Galaxy?}
%%%%
Watson (2011) compared the HI column density with the extinction
with data from the LAB HI survey (Kalberla et al.\ 2005)
and the extinction map of Schlegel et al.\ (1998).
He found that the ${\NH}/{\AV}$ ratio seems to
correlate with $\NH$ toward the Galactic centre.
To investigate the variation of $\NH$/$\AV$ within our Galaxy,
we plot the $\NH/\AV$ ratio distribution projected on
the Galactic longitude-latitude ($l$-$b$) plane
and the Galactic disk in Figure 4.
There is no clear indication for a large-scale variation
of $\NH/\AV$ within the Galaxy.
In Figure~5 we show the $\NH/\AV$ ratio
against (i) $R_g$, the distance away from the
Galactic centre, (ii) $z$, the distance away
from the Galactic plane, (iii) $\nH$, the average
line-of-sight number density of hydrogen,
and (iv) $\NH$. It is apparent that the $\NH/\AV$ ratio
does not show any systematic variation with $R_g$,
$z$, $\nH$ or $\NH$.
We also investigate the $\NH$--$\AV$ relation
for the inner Galaxy (the Galactic first/fourth quadrants)
and outer Galaxy (the second/third quadrants) separately.
We derive
%%%% Equation 8 %%%%
\begin{equation}
{\NH} (\rmHAL\cm^{-2}) = \left(2.04{\pm}0.05\right){\times}10^{21}\,\AV (\magniAL)
\end{equation}
%%%% Equation 8 %%%%
%
for the first/fouth quadrants, and
%
%%%% Equation 9 %%%%
\begin{equation}
{\NH} (\rmHAL\cm^{-2}) = \left(2.09{\pm}0.03\right){\times}10^{21}\,\AV (\magniAL)
\end{equation}
%%%% Equation 9 %%%%
%
for the second/third quadrants (see Figure 6).
They are consistent within each other.
In summary, we can safely conclude that
the X-ray derived $\NH$/$\AV$ ratio
is roughly a constant across the entire Galaxy.

We should note that
it has been well known that the abundances
of heavy elements in our Galaxy
increase toward the Galactic centre
(e.g., see Rolleston et al.\ 2000).
Since AG89 assumed a uniform abundance
for the ISM, at the Galactic centre $\NH$ could
have been overestimated
while at the outer Galaxy $\NH$ could have
been underestimated.
However, it is also well recognized that
the dust grains in the Galactic centre
could be much larger than that of
the diffuse ISM (e.g., see Gao et al.\ 2010,
Shao et al.\ 2017) and the visual extinction
on a per unit mass basis could be smaller.
Therefore, even the Galactic centre is rich
in dust-forming heavy elements and
more heavy elements have been tied up
in dust grains one does not necessarily
expect a lower $\NH/\AV$.
%

%%%%
\subsection{The Gas-to-dust mass ratio}
%%%%
From the hydrogen-to-extinction ratio we can determine
$\Mgas/\Mdust$, the gas-to-dust mass ratio:
%%%% Equation 10 %%%%
\begin{equation}
\frac{\Mgas}{\Mdust}
= 1.086\times1.4\,\muH\,\kappaV\,\left(\frac{\NH}{\AV}\right) ~~,
\end{equation}
%%%% Equation 10 %%%%
%
where $\muH$ is the atomic mass of hydrogen
and $\kappaV$ is the $V$-band
mass extinction coefficient
(i.e., extinction cross section per dust mass).
The coefficient, 1.4, arises from
the contribution of helium.
Taking $\kappaV$ from
Li \& Draine (2001)
and Draine \& Li (2007),
we obtain $\Mgas/\Mdust\approx140$.
%%%% Table 6 %%%%
\begin{table*}
\caption{Hydrogen column density $\NH$ and distance $D$
         compiled from the literature.
         }
\begin{minipage}[b]{1.0\textwidth}
\begin{center}
\begin{tabular}{@{}C{2.8cm}C{1.6cm}C{1.6cm}C{1.6cm}C{1.6cm}C{1.6cm}@{}}
\hline
\hline
      Objects &   Gl &         Gb
              & ${\NH}$ &       $D$ &        References \\
      Name &   (degree) &   (degree)
      & ($10^{21}\rmH\cm^{-2}$) &      (kpc) &            \\
\hline
  G1.0-0.1 &       1.0  &      -0.1  &         75 &       8.5  &        1,2 \\
  G1.9+0.3 &       1.9  &       0.3  &         58 &       8.5  &        3,4 \\
  G5.4-1.2 &       5.4  &      -1.2  &         35 &       5.2  &        1,5 \\
  G8.7-0.1 &       8.7  &      -0.1  &         12 &       4.5  &          6 \\
 G12.0-0.1 &      12.0  &      -0.1  &         49 &      10.9  &        7,8 \\
 G12.8-0.0 &      12.8  &       0.0  &        100 &       4.8  &        7,9 \\
 G15.9+0.2 &      15.9  &       0.2  &         39 &       8.5  &       7,10 \\
 G16.7+0.1 &      16.7  &       0.1  &       47.4 &      14.0  &      11,12 \\
 G18.9-1.1 &      18.9  &      -1.1  &        8.3 &       2.0  &      13,14 \\
 G21.5-0.9 &      21.5  &      -0.9  &       22.4 &       4.8  &      15,16 \\
 G21.8-0.6 &      21.8  &      -0.6  &         28 &       5.5  &      15,17 \\
 G27.4+0.0 &      27.4  &       0.0  &         26 &       8.7  &      18,19 \\
 G28.6-0.1 &      28.6  &      -0.1  &         37 &       7.0  &      20,21 \\
 G28.8+1.5 &      28.8  &       1.5  &         20 &       4.0  &         22 \\
 G29.7-0.3 &      29.7  &      -0.3  &         29 &      10.6  &      23-30 \\
 G31.9+0.0 &      31.9  &       0.0  &         28 &       7.9  &      31-36 \\
 G32.4+0.1 &      32.4  &       0.1  &         52 &      17.0  &         37 \\
 G32.8-0.1 &      32.8  &      -0.1  &        8.1 &       4.8  &      38,39 \\
 G33.6+0.1 &      33.6  &       0.1  &         15 &       7.5  &      40-44 \\
 G34.7-0.4 &      34.7  &      -0.4  &         13 &       2.7  &   43,45-48 \\
 G41.1-0.3 &      41.1  &      -0.3  &         32 &      10.3  &      49,50 \\
 G43.3-0.2 &      43.3  &      -0.2  &         50 &      10.0  &      51-55 \\
 G49.2-0.7 &      49.2  &      -0.7  &         17 &       5.2  &      56-61 \\
 G65.7+1.2 &      65.7  &       1.2  &          2 &       0.9  &      62-64 \\
 G76.9+1.0 &      76.9  &       1.0  &         16 &      10.0  &      65,66 \\
 G85.4+0.7 &      85.4  &       0.7  &        8.3 &       3.8  &      67,68 \\
G148.1+0.8 &     148.1  &       0.8  &        4.9 &       1.0  &      69-71 \\
G156.2+5.7 &     156.2  &       5.7  &        3.2 &       2.0  &      72-75 \\
G195.1+4.3 &     195.1  &       4.3  &       0.11 &       0.25 &      71,76 \\
G201.1+8.3 &     201.1  &       8.3  &       0.25 &       0.28 &   71,77-79 \\
G266.2-1.2 &     266.2  &      -1.2  &        3.5 &       0.75 &      80-84 \\
G272.2-3.2 &     272.2  &      -3.2  &         10 &      10.0  &      85-89 \\
G284.0-1.8 &     294.0  &      -1.8  &          5 &       3.0  &      90,91 \\
G284.2-0.4 &     284.2  &      -0.4  &         15 &       9.0  &      92,93 \\
G284.3-1.8 &     284.3  &      -1.8  &        6.6 &       2.9  &      94,95 \\
G287.4+0.6 &     287.4  &       0.6  &          9 &       3.0  &      96,97 \\
G290.1-0.8 &     290.1  &      -0.8  &        7.3 &       7.0  &      98,99 \\
G291.0-0.1 &     291.0  &      -0.1  &        6.7 &       3.5  &    100,101 \\
G292.2-0.5 &     292.2  &      -0.5  &       14.6 &       8.4  &    102-104 \\
G296.7-0.9 &     296.7  &      -0.9  &       12.4 &       9.8  &        105 \\
G296.8-0.3 &     296.8  &      -0.3  &        6.4 &       9.6  &    106,107 \\
G299.2-2.9 &     299.2  &      -2.9  &        3.5 &       5.0  &    108,109 \\
G304.1-0.2 &     304.1  &      -0.2  &         27 &       7.0  &     93,110 \\
G304.6+0.1 &     304.6  &       0.1  &       38.7 &      10.0  & 89,111,112 \\
G306.3-0.9 &     306.3  &      -0.9  &       19.5 &       8.0  &        113 \\
G308.3-1.4 &     308.3  &      -1.4  &         10 &       9.8  &    114,115 \\
G309.2-0.6 &     309.2  &      -0.6  &        6.5 &       4.0  &        116 \\
G309.8-2.6 &     309.8  &      -2.6  &        3.9 &       2.5  &    117,118 \\
G310.6-1.6 &     310.6  &      -1.6  &       20.9 &      10.0  &     93,119 \\
G311.5-0.3 &     311.5  &      -0.3  &       27.7 &      13.8  & 89,112,120 \\
G313.3+0.1 &     313.3  &       0.1  &       18.5 &       3.0  &     93,121 \\
G313.6+0.3 &     313.6  &       0.3  &       35.4 &       5.6  &    121-123 \\
G326.3-1.8 &     326.3  &      -1.8  &        3.5 &       5.1  &    124,125 \\
G327.1-1.1 &     327.1  &      -1.1  &       18.8 &       9.0  &    126-128 \\
G327.4+0.4 &     327.4  &       0.4  &         24 &       4.3  &    129,130 \\
G330.2+1.0 &     330.2  &       1.0  &         26 &       7.8  & 89,130,131 \\
G337.2+0.1 &     337.2  &       0.1  &         83 &      11.0  &    132-134 \\
G337.2-0.7 &     337.2  &      -0.7  &       32.3 &       9.0  &     89,135 \\
G337.8-0.1 &     337.8  &      -0.1  &       67.5 &      12.3  &    136,137 \\
G338.3-0.0 &     338.3  &       0.0  &        150 &      12.8  &    138,139 \\
G343.7-2.3 &     343.7  &      -2.3  &          5 &       2.6  & 71,140,141 \\
G344.7-0.1 &     344.7  &      -0.1  &         45 &       6.3  &        142 \\
G346.6-0.2 &     346.6  &      -0.2  &       22.7 &       11.0 & 120,137,143-146 \\
\end{tabular}
\end{center}
\end{minipage}
\end{table*}

\begin{table*}
\begin{minipage}[b]{1.0\textwidth}
\begin{center}
\begin{tabular}{@{}C{2.8cm}C{1.6cm}C{1.6cm}C{1.6cm}C{1.6cm}C{1.6cm}@{}}
G347.3-0.5 &     347.3  &      -0.5  &        7.8 &       1.0  &     146-149 \\
G348.5+0.1 &     348.5  &       0.1  &         35 &       7.9  & 120,150,151 \\
G348.7+0.3 &     348.7  &       0.3  &         38 &      13.2  &     151-153 \\
G349.7+0.2 &     349.7  &       0.2  &         64 &      11.5  &     154,155 \\
G350.1-0.3 &     350.1  &      -0.3  &         33 &       9.0  &        154  \\
G352.7-0.1 &     352.7  &      -0.1  &         32 &       7.6  &     156,157 \\
G353.6-0.7 &     353.6  &      -0.7  &       18.2 &       3.2  &     158-160 \\
G355.6-0.0 &     355.6  &       0.0  &         55 &      13.0  &     145,161 \\
G357.7-0.1 &     357.7  &      -0.1  &        100 &      11.8  &      31,162 \\
G359.0-0.9 &     359.0  &      -0.9  &         16 &       3.7  &  89,163,164 \\
G359.1-0.5 &     359.0  &      -0.5  &         25 &       7.6  &     163-165 \\
\hline
\end{tabular}
\end{center}

References: (1) Frail (2011);
(2) Nobukawa et al.\ (2009);
(3) Reynolds et al.\ (2008);
(4) Reynolds et al.\ (2009);
(5) Kaspi et al.\ (2001);
(6) Hewitt \& Yusef-Zadeh (2009);
(7) Kilpatrick et al.\ (2016);
(8) Yamauchi et al.\ (2014a);
(9) Halpern et al.\ (2012);
(10) Reynolds et al.\ (2006);
(11) Tian et al.\ (2017, in preparation)
(12) Helfand et al.\ (2003a);
(13) Kassim et al.\ (1994);
(14) Harrus et al.\ (2004);
(15) Tian \& Leahy (2008a);
(16) Safi-Harb et al.\ (2001);
(17) Bocchino et al.\ (2012);
(18) Tian \& Leahy (2008b);
(19) Kumar et al.\ (2014);
(20) Bamba et al.\ (2003);
(21) Ueno et al.\ (2003);
(22) Misanovic et al.\ (2010);
(23) Su et al.\ (2009);
(24) Temim et al.\ (2012);
(25) Leahy \& Tian (2008);
(26) Martin et al.\ (2014);
(27) Blanton \& Helfand (1996);
(28) Helfand et al.\ (2003b);
(29) Morton et al.\ (2007);
(30) Ng et al.\ (2008);
(31) Frail et al.\ (1996);
(32) Rho \& Petre (1998);
(33) Radhakrishnan et al.\ (1972);
(34) Rho \& Petre (1996);
(35) Sugizaki et al.\ (2001);
(36) Chen \& Slane (2001);
(37) Yamaguchi et al.\ (2004);
(38) Zhou \& Chen (2011);
(39) Bamba et al.\ (2016);
(40) Auchettl et al.\ (2014);
(41) Giacani et al.\ (2009);
(42) Sato et al.\ (2016);
(43) Rho \& Petre (1998);
(44) Tsunemi \& Enoguchi (2002);
(45) Park et al.\ (2013);
(46) Harrus et al.\ (1997);
(47) Cox et al.\ (1999);
(48) Uchida et al.\ (2012a);
(49) Jiang et al.\ (2010);
(50) Yamaguchi et al.\ (2014);
(51) Safi-Harb et al.\ (2000);
(52) Moffett \& Reynolds (1994);
(53) Hwang et al.\ (2000);
(54) Brogan \& Troland (2001);
(55) Keohane et al.\ (2007);
(56) Green et al.\ (1997);
(57) Koo et al.\ (2002);
(58) Tian \& Leahy (2013);
(59) Koo et al. (2005);
(60) Hanabata et al.\ (2013);
(61) Sasaki et al.\ (2014);
(62) Kothes et al.\ (2004);
(63) Arzoumanian et al.\ (2004);
(64) Arzoumanian et al.\ (2008);
(65) Marthi et al.\ (2011);
(66) Arzoumanian et al.\ (2011);
(67) Kothes et al.\ (2001);
(68) Jackson et al.\ (2008);
(69) McGowan et al.\ (2006);
(70) Tepedelenli{\v g}lu \& {\"O}gelman (2007);
(71) Verbiest et al.\ (2012);
(72) Katsuda et al.\ (2009);
(73) Uchida et al.\ (2012b);
(74) Kassim et al.\ (1994);
(75) Reich et al.\ (1992);
(76) Halpern \& Wang (1997);
(77) De Luca et al.\ (2005);
(78) Marshall \& Schulz (2002);
(79) Greiveldinger et al.\ (1996);
(80) Kargaltsev et al.\ (2002);
(81) Bamba et al.\ (2005);
(82)Pannuti et al.\ (2010a);
(83) Hiraga et al.\ (2009);
(84) Allen et al.\ (2015);
(85) Sanchez-Ayaso et al.\ (2013);
(86) Kamitsukasa et al.\ (2015);
(87) Mcentaffer et al.\ (2013);
(88) Harrus et al.\ (2001);
(89) Pavlovi{\'c}  et al.\ (2013);
(90) Camilo et al.\ (2004a);
(91) Kargaltsev \& Pavlov (2010);
(92) Saz Parkinson et al.\ (2010);
(93) Kargaltsevet al.\ (2013);
(94) Abramowski et al.\ (2012a);
(95) Ruiz \& May (1986);
(96) Gonzalez et al.\ (2006);
(97) Abdo et al.\ (2009);
(98) Kamitsukasa et al.\ (2016);
(99) Reynoso et al.\ (2006);
(100) Slane et al.\ (2012);
(101) Roger et al.\ (1986);
(102) Kumar et al.\ (2012);
(103) Pivovaroff et al.\ (2001);
(104) Caswell et al.\ (2004);
(105) Prinz \& Becker (2013);
(106) Sanchez-Ayaso et al.\ (2012);
(107) Gaensler et al.\ (1998);
(108) Park et al.\ (2007);
(109) Slane et al.\ (1996);
(110) Abramowski et al.\ (2012b);
(111) Gelfand et al.\ (2013);
(112) Caswell et al.\ (1975);
(113) Reynolds et al.\ (2013);
(114) Hui et al.\ (2012);
(115) Prinz \& Becker (2012);
(116) Rakowski et al.\ (2001);
(117) Lemoine-Goumard et al.\ (2011);
(118) Camilo et al.\ (2004b);
(119) Renaud et al.\ (2010);
(120) Pannuti et al.\ (2014);
(121) Ng et al.\ (2005);
(122) Roberts et al.\ (2001);
(123) Van Etten \& Romani (2010);
(124) Yatsu et al.\ (2013);
(125) Rosado et al.\ (1996);
(126) Temim et al.\ (2009);
(127) Temim et al.\ (2015);
(128) Bocchino \& Bandiera (2003);
(129) Enoguchi et al.\ (2002);
(130) McClure-Griffiths et al.\ (2001);
(131) Torii et al.\ (2006);
(132) Esposito et al.\ (2009);
(133) Woods et al.\ (1999);
(134) Sarma et al.\ (1997);
(135) Rakowski et al.\ (2006);
(136) Zhang et al.\ (2015);
(137) Koralesky et al.\ (1998);
(138) Gotthelf et al.\ (2014);
(139) Lemiere et al.\ (2009);
(140) Romani et al.\ (2005);
(141) Gotthelf et al.\ (2002);
(142) Giacani et al.\ (2011);
(143) Pannuti et al.\ (2014);
(144) Yamauchi et al.\ (2013);
(145) Yamauchi et al.\ (2008);
(146) Acero et al.\ (2009);
(147) Takahashi et al.\ (2008);
(148) Tanaka et al.\ (2008);
(149) Fukui et al.\ (2003);
(150) Yamauchi et al.\ (2014b);
(151) Tian \& Leahy (2012);
(152) Nakamura et al.\ (2009);
(153) Aharonian et al.\ (2008a);
(154) Yasumi et al.\ (2014);
(155) Tian \& Leahy (2014);
(156) Giacani et al.\ (2009);
(157) Sezer \& G{\"o}k (2014);
(158) Tian et al.\ (2010);
(159) Bamba et al.\ (2012);
(160) Tian et al.\ (2008c);
(161) Minami et al.\ (2013);
(162) Gaensler et al.\ (2003);
(163) Bamba et al.\ (2000);
(164) Bamba et al.\ (2009);
(165) Aharonian et al.\ (2008b)

%\end{minipage}
%\end{multicols}
\end{minipage}
\end{table*}
%%%% Table 6 %%%%

%%%% Figure 7 %%%%
\begin{figure}
\centering
\includegraphics[width=0.48\textwidth, angle=0]{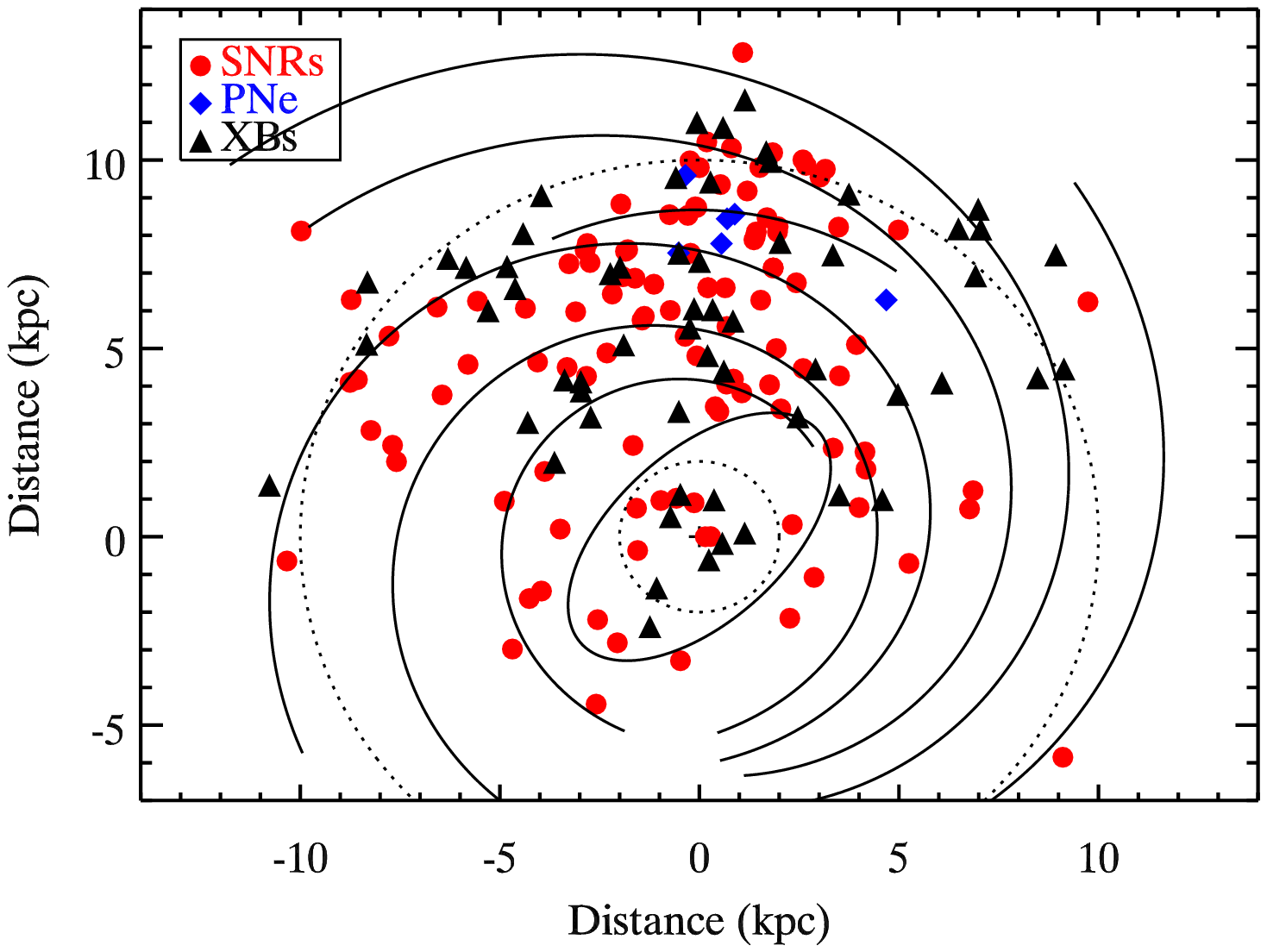}
\caption{The projection distribution of the AG89 sample on the Galactic plane.
         The Galactic center is marked by a plus sign.
         The dot circles and solid lines are the same as the circles and lines in Figure~4.
         }
\label{fig7}
\end{figure}
%%%% Figure 7 %%%%

%%%% Figure 8 %%%%
\begin{figure}
\centering
\includegraphics[width=0.48\textwidth, angle=0]{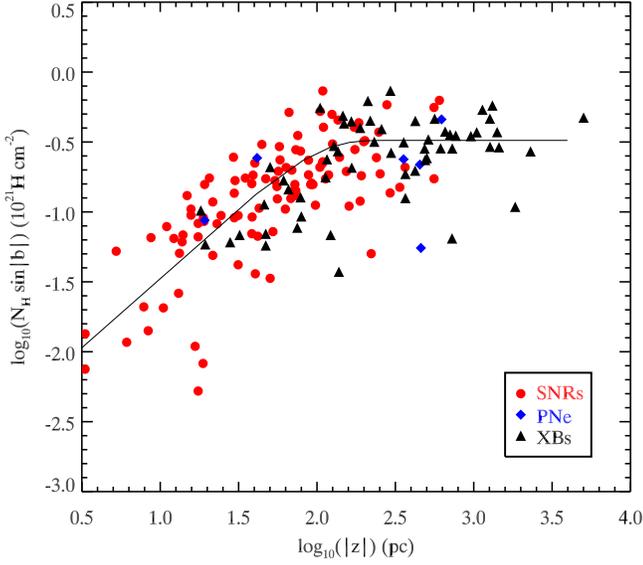}
\caption{Logarithm of $\NH\sin|b|$ vs. the logarithm of $z$
              (distance above the Galactic plane).
              The solid line is the best fit to the data
              with a midplane density of
              $\nH(0) = 1.11\pm0.15\cm^{-3}$
              and a scale hight of $h = 75.5\pm12.4\pc$.
         }
\label{fig8}
\end{figure}
%%%% Figure 8 %%%%

%%%% Figure 9 %%%%
\begin{figure}
\centering
\includegraphics[width=0.48\textwidth, angle=0]{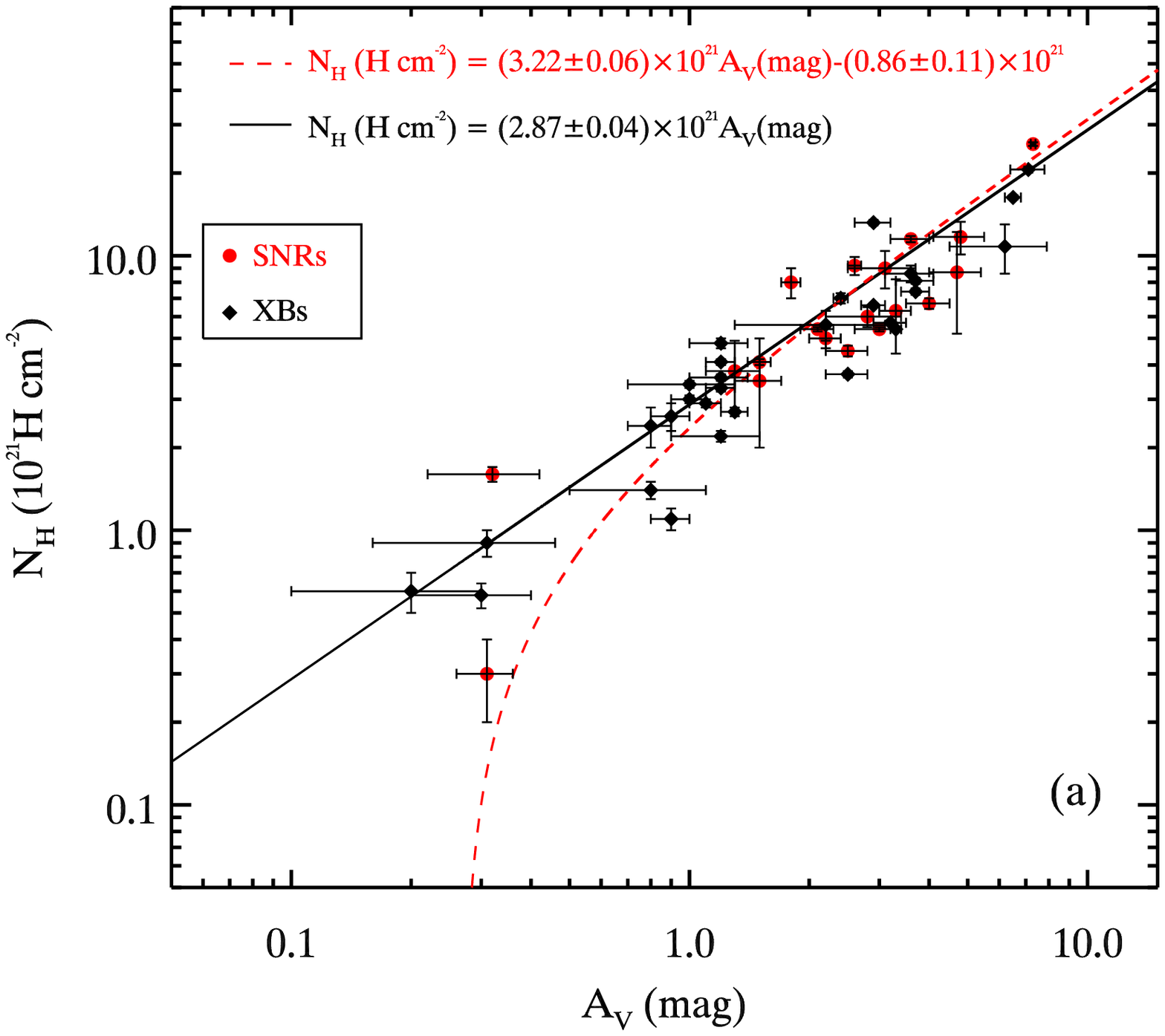}
\includegraphics[width=0.48\textwidth, angle=0]{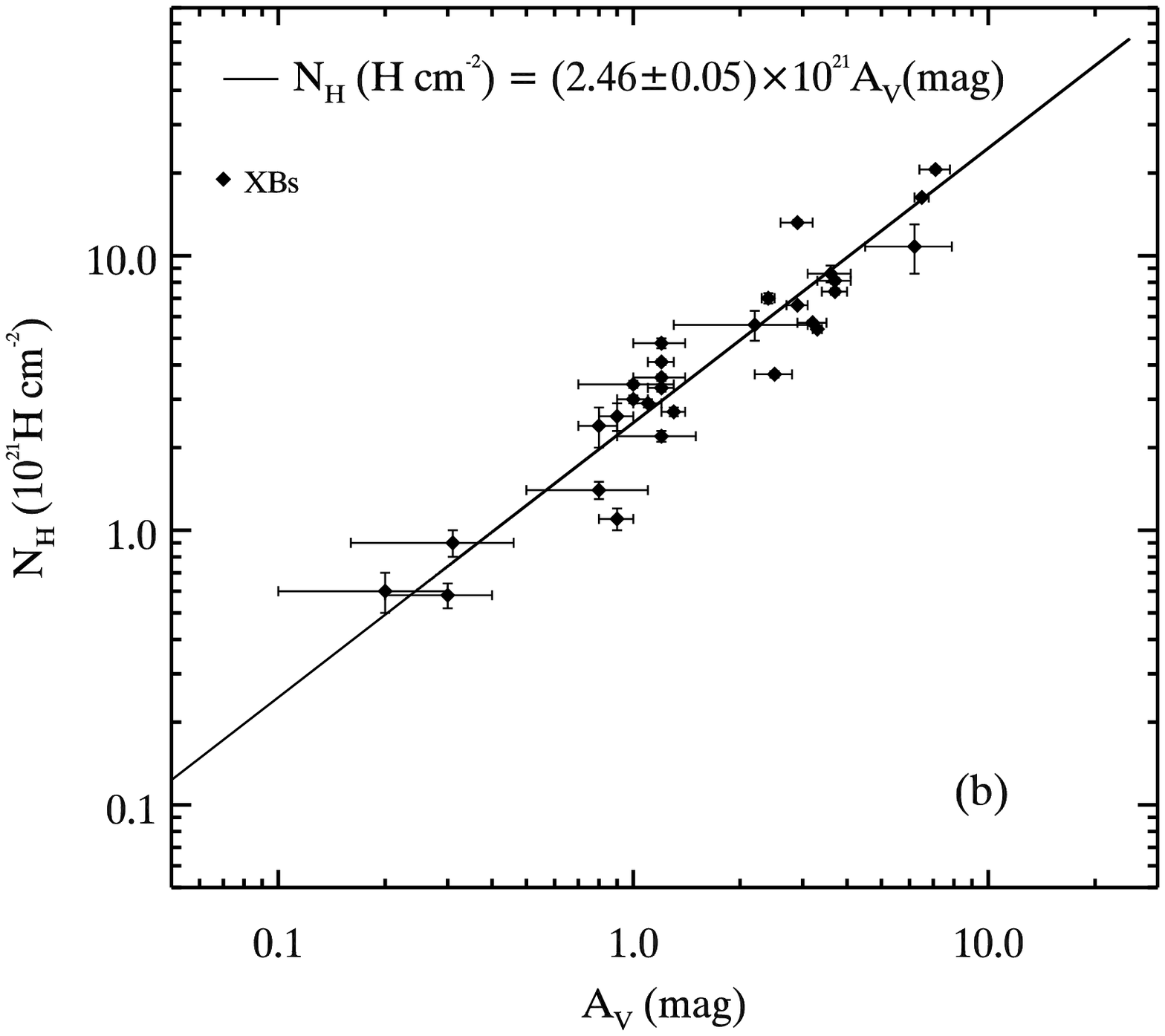}
\includegraphics[width=0.48\textwidth, angle=0]{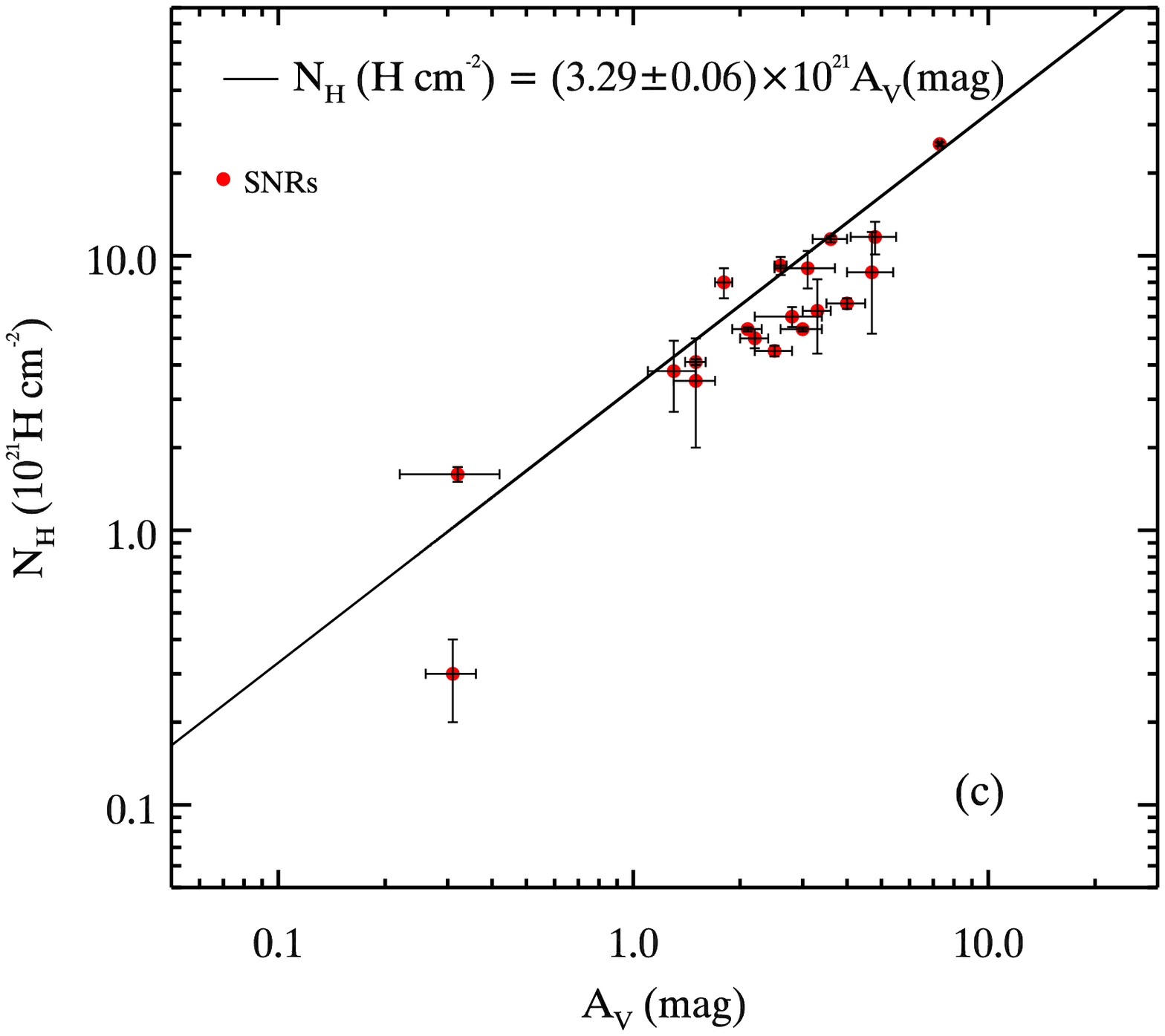}
\caption{Top (a): $\NH$ against $\AV$ for the whole W00 sample.
              Middle (b): $\NH$ against $\AV$ for the XBs of the W00 sample.
              Bottom (c): $\NH$ against $\AV$ for the SNRs of the W00 sample.
              }
\label{fig9}
\end{figure}
%%%% Figure 9 %%%%
%%%%
\subsection{The gas distribution of the Galaxy}
%%%%
When both $\NH$ and $D$
(the distance of the source to Earth) are known,
we are able to study the distribution
of the gas in the Galaxy.
To this end, we enlarge the sample in Table 1
by including 74 SNRs with known $\NH$ (AG89 abundance)
and $D$ from the literature (see Table 6).
The final sample includes 175 calibrators.
Figure 7 shows the projection distribution
of the sample on the Galactic plane.

The distribution of hydrogen away from the Galactic plane is
usually described by one to three Gaussian and/or exponential functions,
(e.g., see Bohlin, Savage \& Drake 1978, Diplas \& Savage 1994,
Dickey \& Lockman 1990, Kalberla \& Kerp 2009).
A commonly-used HI distribution between the Galactic radius
4\,kpc and 8\,kpc is given by Dickey \& Lockman (1990):
$\nH(z) = n_{\rm H,1}(0)\,\exp\left(-z^2/2h_1^2\right)
+ n_{\rm H,2}(0)\,\exp\left(-z^2/2h_2^2\right)
+ n_{\rm H,3}(0)\,\exp\left(-z^2/2h_3^2\right)$,
where the midplane densities are
$n_{\rm H,1}(0)\approx0.395\cm^{-3}$,
$n_{\rm H,2}(0)\approx0.107\cm^{-3}$, and
$n_{\rm H,3}(0)\approx0.064\cm^{-3}$,
and the scale heights
are $h_1\approx90\pc$,
$h_2\approx225\pc$, and
$h_3\approx403\pc$.
A more recent study for the Galactic radius
between 5\,kpc and 35\,kpc suggested that
the HI scale hight varies with the distance away
from the Galactic centre:
$h(R_g)\approx150\,\exp\left\{(R_g-R_0)/9800\right\}\pc$,
where $R_0$ is the Galactic radius of the Sun
(Kalberla \& Kerp 2009).
Figure 8 shows $\NH\sin(|b|)$ versus $z$.
To avoid the influence of the Galactic bulge and wrap,
we only make use of the data within $2\kpc\simlt R_g \simlt10\kpc$.
We use a one-component Gaussian model,
$\nH(z) = \nH(0)\,\exp\left(-z^2/2h^2\right)$,
to fit the data. The best-fit is given by
$\nH(0) = 1.11\pm0.15\cm^{-3}$
and $h = 75.5\pm12.4\pc$
(see Figure~8).
Compared with that of Dickey \& Lockman (1990),
our results suggest a larger midplane density
and a sightly smaller scale height.
This might be caused by the fact that the $\NH$
column densities adopted here are based on
X-ray absorption data and contain the contribution of molecular and ionized ISM,
while the 21\,cm emission only presents
atomic hydrogen. We are not able to estimate the scale
heights of any other extended components,
because the majority of our sample has $z<1000\pc$.
%

%%%% Figure 10 %%%%
\begin{figure}
\centering
\includegraphics[width=0.48\textwidth, angle=0]{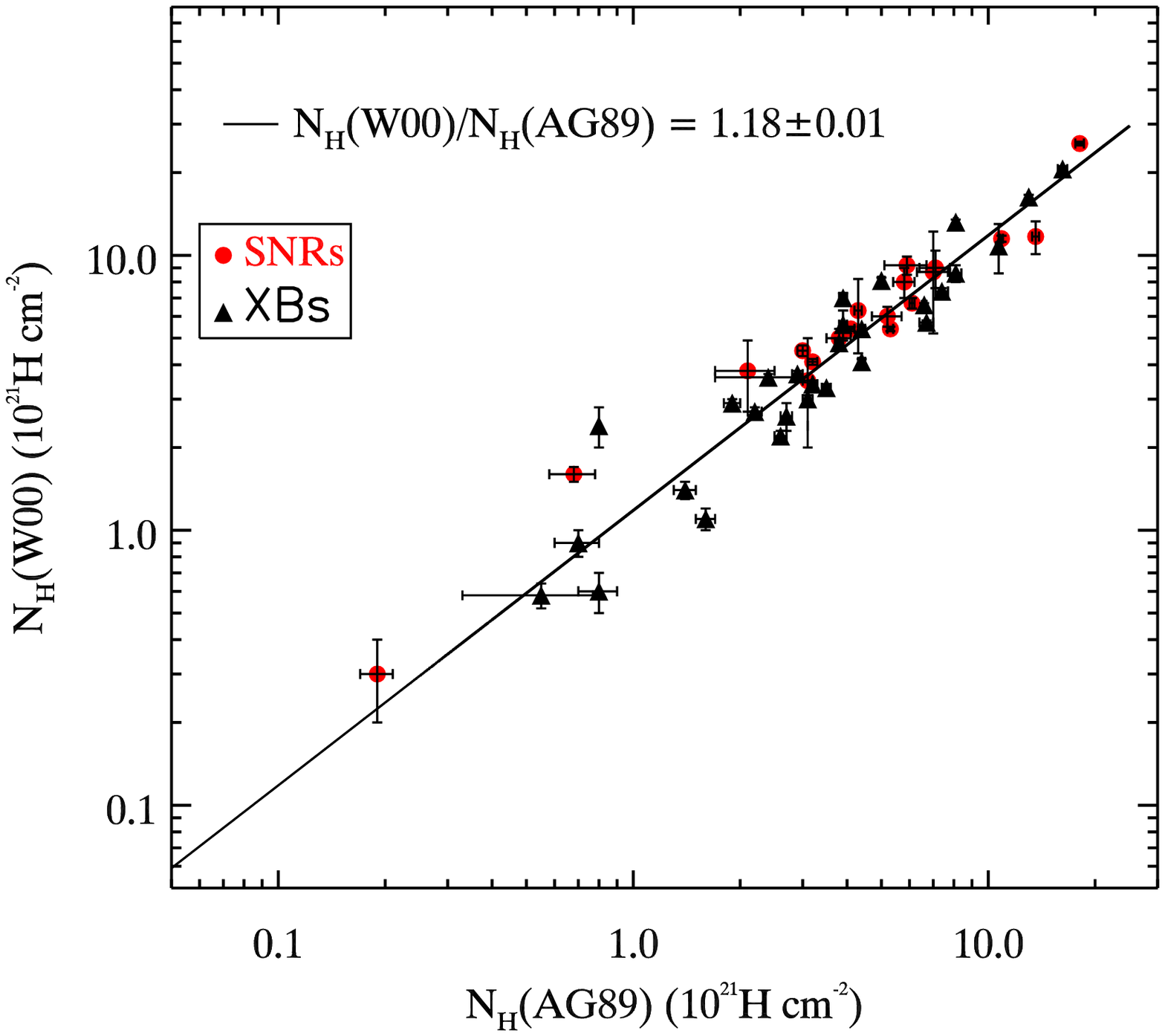}
\caption{$\NH$ derived from subsolar abundances (W00)
               against $\NH$ derived from solar abundances (AG89).
               }
\label{fig10}
\end{figure}
%%%% Figure 10 %%%%

The neutral hydrogen surface density
is nearly constant within the Galactic
radius of $\simali$12.5\,kpc (Kalberla \& Kerp 2009).
A mean surface density of $\simali$5$\Msun\pc^{-2}$
was given by Dickey \& Lockman (1990).
More recently, Kalberla \& Dedes (2008)
argued for a surface density two times
larger than that of Dickey \& Lockman (1990).
Based on the best-fit $\nH(0)$ and $h$ values,
here we estimate a mean gas surface density
of the Galaxy in the Galactic radius range
of $\simali$2\,kpc to $\simali$10\,kpc
to be $\sum_{\rm gas}\simali$1.4\,$\nH(0)\times2.4\,h
\approx7.0\Msun\pc^{-2}$.

Due to the limited amount of data,
we are not able to accurately determine
the radial distribution of the midplane density,
but only give an rough estimation
of $\nH\approx4.33\pm12.87\cm^{-3}$
for $2.0\kpc < R_g < 4.5\kpc$,
$\nH\approx1.15\pm5.35\cm^{-3}$
for $4.5\kpc < R_g < 6.5\kpc$,
$\nH\approx1.18\pm0.77\cm^{-3}$
for $6.5\kpc < R_g < 8.5\kpc$, and
$\nH\approx1.13\pm0.81\cm^{-3}$
for $8.5\kpc < R_g < 10\kpc$.
All regions are limited to $|z|<75\pc$.
%

%%%%
\subsection{$\NH/\AV$ from the W00 Sample}
%%%%
Since the X-ray-derived $\NH$ is sensitive
to the abundances of heavy elements
(relative to H),
one could expect a higher $\NH/\AV$
for the W00 sources of which $\NH$ is determined
with subsolar abundances
compared to that of the AG89 sources
of which $\NH$ is determined
with solar abundances.
By fitting the whole W00 sample
which include both SNRs and XBs,
we indeed obtain a larger
$\langle\NH/\AV\rangle$:
%
%%%% Equation 11 %%%%
\begin{equation}
\NH (\rmHAL\cm^{-2}) = \left(2.87\pm0.04\right)
\times10^{21}\,\AV (\magniAL) ~~.
\end{equation}
%%%% Equation 11 %%%%
We also obtain
%%%% Equation 12 %%%%
\begin{equation}
{\NH} ({\rmHAL}{\cm}^{-2}) = \left(3.30{\pm}0.06\right)
{\times}10^{21}\,\AV ({\magniAL})
\end{equation}
%%%% Equation 12 %%%%
for the SNRs of the W00 sample,
and
%%%% Equation 13 %%%%
\begin{equation}
{\NH} ({\rmHAL}{\cm}^{-2}) = \left(2.46{\pm}0.05\right)
{\times}10^{21}\,{\AV} ({\magniAL}) ~~
\end{equation}
%%%% Equation 13 %%%%
for the XBs of the W00 sample.
Again, the mean $\langle\NH/\AV\rangle$ ratio
from SNRs differs from that of XBs by $\simali$34\%.
If we exclude three SNRs
(G54.1+0.3, G116.9+0.2, and G184.6-5.8)
which appear to be outliers,
we derive
%
%%%% Equation 14 %%%%
\begin{equation}
{\NH} ({\rmHAL}{\cm}^{-2}) = \left(2.51{\pm}0.10\right)
{\times}10^{21}\,\AV ({\magniAL})
\end{equation}
%%%% Equation 14 %%%%
%
for SNRs, and
%%%% Equation 15 %%%%
\begin{equation}
{\NH} ({\rmHAL}{\cm}^{-2}) = \left(2.47{\pm}0.04\right)
{\times}10^{21}\,{\AV} ({\magniAL})
\end{equation}
%%%% Equation 15 %%%%
%
for the whole W00 sample.
Similar to the AG89 sample,
with the three SNR outliers excluded,
the $\langle\NH/\AV\rangle$ of SNRs
agree with that of XBs.
We tabulate the best-fit results in Table~5.

The $\langle\NH/\AV\rangle$ derived here
is smaller by $\simali$15\% than that of
Foight et al.\ (2016) who assumed
subsolar abundances for 17 SNRs
(see Table~3)
and appreciably larger than most of
those derived from HI and/or H${_2}$ UV absorption
and HI 21\,cm emission (see Tables~1,2).
Two factors might be responsible for
the inconsistency.
First, our sample is larger than previous ones
and it includes 29 XBs which do not suffer
the problem with which SNRs may be confronted
(i.e., the measured $\NH$ and $\AV$
may not be for the same region.
Second, the $\NH$ column derived from
the UV absorption data of HI and/or H${_2}$
and the HI 21\,cm emission data
only counts neutral and/or molecular hydrogen,
while the $\NH$ value determined from
the X-ray absorption data includes
atomic, molecular and ionic hydrogen.

With
$\langle\NH/\AV\rangle\approx2.5\times10^{21}\magni\cm^2\rmH^{-1}$,
we estimate the gas-to-dust mass ratio to be
$\Mgas/\Mdust\approx170$, again, based on
the mass extinction coefficient of
Li \& Draine (2001) and Draine \& Li (2007).
As expected, the gas-to-dust mass ratio
of the W00 sample exceeds that of the AG89 sample
by $\simali$20\%.
In Figure~10, we plot $\NH$ from the W00 sample
against that from the AG89 sample.
The fact that they are closely related
with $\NH({\rm W00})/\NH({\rm W00})\approx1.2$
suggests that in future studies one may simply scale
$\NH$ derived from subsolar abundances
by a factor of $\simali$1.2
when converting to $\NH$
derived from solar abundances.
%

%%%%
\subsection{Caveats}
%%%%
Even though in our analysis
we have tried our best to avoid
systematic uncertainties
(e.g., for SNRs adopting $\NH$ and $\AV$
determined from the same region if possible,
excluding those XBs with intrinsic
local X-ray absorptions),
there may still be sources
of potential uncertainty:
\begin{itemize}
\item Since we compile $\NH$ and $\AV$
from the literature,
for some SNRs the adopted $\NH$ and $\AV$
may not be for the same region
(e.g., SNR G132.7).
\item Since a number of SNRs only
have one or two measures
of $\NH$ and/or $\AV$, we are not able to
take into account the uncertainties caused
by spatial variations.
\item For most SNRs,
$\AV$ was measured from line ratios,
e.g., H$\alpha$(6563$\Angstrom$)/H$\beta$(4861$\Angstrom$),
[SII]($\sim$10320$\Angstrom$)/[SII]($\sim$4068$\Angstrom$), and
[FII]($\sim$16435$\Angstrom$)/[FII]($\sim$12567$\Angstrom$).
Taking the Balmer decrement as an example,
the intrinsic ratio of H$\alpha$/H$\beta$
is commonly taken to be $\simali$3
under the ``Case A'' condition,
in which the nebula is always optically thin
even for Lyman-line photons.
However, this may not be always true for SNRs.
Previous studies have showed that H$\alpha$/H$\beta$
could vary from $\simali$2.9 to $\simali$4.2
and even up to $\simali$6 (e.g., see Raymond 1979,
Shull \& McKee 1979, Ghavamian et al.\ 2001).
\item For some XBs, the optical extinction was estimated
by comparing the observed colour with the intrinsic colour
derived from the stellar spectral type.
However, if the optical flux contribution of
the accretion disk or the absorption of
the stellar wind is not negligible,
the derived spectral type of the primary star
may be incorrect and this may result in
an incorrect $\AV$ (e.g., for EXO 0748-676,
Schoembs \& Zoeschinger (1990) estimated
$\AV\approx1.3\pm0.1\magni$
based on its stellar spectral type,
while Pearson et al.\ (2006)
estimated $\AV\approx0.48\pm0.26\magni$
from the widths of the sodium doublet absorption lines).
\end{itemize}
%

%%%%
\section{Summary}
%%%%
The $\NH$--$\AV$ relation is explored based on
the X-ray absorption data of a large AG89 sample of
Galactic lines of sight toward 35 supernova remnants,
6 planetary nebulae, 70 X-ray binaries
for which $\NH$ was derived with solar abundances,
as well as a smaller W00 sample of
19 supernova remnants and 29 X-ray binaries
for which $\NH$ was derived with subsolar abundances.
For the AG89 sample,
our principal results are as follows:
\begin{enumerate}
\item We derive a mean hydrogen-to-extinction ratio
          of $\langle\NH/\AV\rangle = \left(2.08\pm0.02\right)
          \times10^{21}\rmH\cm^{-2}\magni^{-1}$
          for the whole Galaxy.
\item The X-ray-derived $\NH/\AV$ ratio
          is generally constant across the Galaxy.
          It does not appear to correlate with
          $n_{\rm H}$ (the line of sight mean
          number density of hydrogen),
          $R_g$ (the distance away from the Galactic centre),
          and $z$ (the distance above or below the Galactic plane).
          The $\langle\NH/\AV\rangle$ ratio derived for
          the 1st and 4th galactic quadrants
          ($\langle\NH/\AV\rangle = \left(2.04\pm0.05\right)
          \times10^{21}\rmH\cm^{-2}\magni^{-1}$)
          is consistent with that of the 2nd and 3rd
          galactic quadrants
          ($\langle\NH/\AV\rangle = \left(2.09\pm0.03\right)
           \times10^{21}\rmH\cm^{-2}\magni^{-1}$).
\item We estimate a gas-to-dust mass ratio
          of $\Mgas/\Mdust\approx140$
          from $\langle\NH/\AV\rangle = 2.08
          \times10^{21}\rmH\cm^{-2}\magni^{-1}$.
\item The distribution of hydrogen in the Galaxy
          is explored with the addition of an additional
          sample of 74 supernova remnants
          for which both $\NH$ and distances are known.
          It is found that, between the Galactic radius of
          2\,kpc to 10\,kpc, the vertical distribution of hydrogen
          can be roughly described by a Gaussian function
          with a mid-plane density of
          $n_{\rm H}(0) = 1.11\pm0.15\cm^{-3}$ and
          a scale hight of $h = 75.5\pm12.4\pc$.
          We also estimate a total gas surface density of
          $\sum_{\rm gas}\sim7.0\Msun\pc^{-2}$.
\end{enumerate}

Similarly, for the W00 sample we derive
$\NH/\AV = \left(2.47\pm0.04\right)
\times10^{21}\rmH\cm^{-2}\magni^{-1}$
for the whole Galaxy which is higher by
$\simali$20\% than that of the AG89 sample.
We also find that $\NH$ derived from subsolar
abundances exceeds that from solar abundances
by $\simali$20\%, suggesting that
in future studies one may simply scale
$\NH$ of subsolar abundances
by a factor of $\simali$1.2
when converting to $\NH$ of solar abundances.
%

%%%%
\section*{Acknowledgements}
%%%%

HZ and WWT acknowledge support from
NSFC (10273025, 11473038, 11603039).
AL is supported in part
by NSF AST-1311804 and NASA NNX14AF68G.
We thank Patrick Slane, Linli Yan,
and the referee Tolga G\"uver
for their very helpful comments and suggestions
which substantially improve the quality of our paper.
We also thank H.~Su, S.~Yao and D.~Wu for their help
during the preparation of this paper.

%%%%
\section*{Appendix: A list of $\NH$, $\AV$ and $D$ data
         compiled from the literature}
%%%%
%%%%

\begin{landscape}
%\begin{table*}
%Table~8. Hydrogen column density $\NH$ and distance $\AV$ compiled from the literature.
\begin{figure}
%\caption{Hydrogen column density $\NH$ and distance $\AV$ compiled from the literature.}
{\tiny
\begin{minipage}[b]{1.0\textwidth}
\begin{center}
\textbf{Table 7.} Hydrogen column density $\NH$ and distance $D$ compiled from the literature.
\begin{supertabular}{@{}C{1.2cm}C{0.25cm}C{0.25cm}C{2.4cm}C{0.8cm}C{2.4cm}C{0.8cm}C{2.4cm}C{0.8cm}C{2.4cm}C{0.8cm}C{0.8cm}C{0.8cm}C{2.4cm}C{0.8cm}C{0.8cm}@{}}
\hline
\hline

    Object & Type  & $D$   & References & $\NH({\rm AG89})$ & References
    & Accept ``A'' & $\NH({\rm AG89})$ & $\NH({\rm W00})$ & References
    & Accept ``A''  & $\NH({\rm W00})$ & $A_V$ & References
    & Accept ``A''  & $A_V$ (mag) \\
          &       & (kpc) & for $D$ & (${10}^{21}\,{\rm H}\,{\rm
            cm}^{-2}$) & for $\NH({\rm AG89})$ & or Reject ``R'' &
          This Work & (${10}^{21}\,{\rm H}\,{\rm cm}^{-2}$) & for
          $\NH({\rm W00})$ & or Reject ``R'' & This Work & (mag) & for $A_V$ & or Reject ``R'' & This Work \\
\hline
    G4.5+6.8 & SNR   & 5.1   & Sankrit et al.\ 2016 & 5.3$\pm$0.4 & Bamba et al.\ 2005 & A     & 5.3$\pm$0.1 & 5.4$\pm$0.1 & Foight et al.\ 2016 & A     & 5.4$\pm$0.1 & 2.8$\pm$0.3 & Blair et al.\ 1991 & A     & 3.0$\pm$0.4 \\
    north &       &       &       & 3.5$\pm$0.2 & Cassam-Chenai et al.\ 2004 & A     &       &       &       &       &       & 2.5$\pm$0.9   & Oliva et al.\ 1989 & A     &  \\
          &       &       &       & 5.7$\pm$0.1 & Kinugasa \& Tsunemi 1999 & A     &       &       &       &       &       & 3.5   & Leibowitz \& Danziger 1983 & A     &  \\
          &       &       &       &       &       &       &       &       &       &       &       & 3.3   & Dennefeld 1982 & A     &  \\
          &       &       &       &       &       &       &       &       &       &       &       & 2.2$\pm$0.6 & van den Bergh \& Kamper 1977 & A     &  \\
          &       &       &       &       &       &       &       &       &       &       &       &       &       &       &  \\
\hline

\end{supertabular}
\end{center}
\end{minipage}
}
\end{figure}
%\end{table*}
\end{landscape}

\begin{landscape}
%\begin{table*}
\begin{figure}
%\caption{Hydrogen column density $\NH$ and distance $D$ compiled from the literature.}
{\tiny
\begin{minipage}[b]{1.0\textwidth}
\begin{center}
\begin{supertabular}{@{}C{1.2cm}C{0.25cm}C{0.25cm}C{2.4cm}C{0.8cm}C{2.4cm}C{0.8cm}C{2.4cm}C{0.8cm}C{2.4cm}C{0.8cm}C{0.8cm}C{0.8cm}C{2.4cm}C{0.8cm}C{0.8cm}@{}}
\hline

    G119.5+10.2 & SNR   & 1.4   & Pineault et al.\ 1993 & 1.8$\pm$0.3 & Caraveo et al.\ 2010 & A     & 2.1$\pm$0.4 & 3.8$\pm$1.1 & Foight et al.\ 2016 & A     & 3.8$\pm$1.1 & 1.27$\pm$0.14 & Mavromatakis et al.\ 2000 & A     & 1.3$\pm$0.2 \\
          &       &       &       & 2.8$\pm$0.6 & Slane et al.\ 1997 & A     &       &       &       &       &       & 1.5   & Fesen et al.\ 1981 & A     &  \\
          &       &       &       & 2.3$\pm$0.6 & Seward et al.\ 1995b & A     &       &       &       &       &       &       &       &       &  \\
          &       &       &       &       &       &       &       &       &       &       &       &       &       &       &  \\
    G120.1+1.4 & SNR   & 3.0   & Zhou et al.\ 2016 & 6.6$\pm$0.4 & Cassam-Chenai et al.\ 2007 & A     & 5.8$\pm$0.4 & 8.0$\pm$1.0 & Foight et al.\ 2016 & A     & 8.0$\pm$1.0 & 1.9$\pm$0.1 & Ruiz-Lapuente 2004 & A     & 1.8$\pm$0.1 \\
          &       &       & Tian \& Leahy 2011 & 6.4$\pm$0.5 & Warren et al.\ 2005 & A     &       &       &       &       &       & $\sim$1.6 & Ghavamian et al.\ 2000 & A     &  \\
          &       &       &       & 4.7$\pm$0.3 & Hwang et al.\ 2002 & A     &       &       &       &       &       & 2.1$\pm$0.5 & Chevalier et al.\ 1980 & A     &  \\
          &       &       &       & 8.3   & Decourchelle et al.\ 2001 & A     &       &       &       &       &       &       &       &       &  \\
\hline
   \end{supertabular}
\end{center}
\end{minipage}
}
\end{figure}
%\end{table*}
\end{landscape}

\begin{landscape}
%\begin{table*}
\begin{figure}
%\caption{Hydrogen column density $\NH$ and distance $D$ compiled from the literature.}
{\tiny
\begin{minipage}[b]{1.0\textwidth}
\begin{center}
\begin{supertabular}{@{}C{1.2cm}C{0.25cm}C{0.25cm}C{2.4cm}C{0.8cm}C{2.4cm}C{0.8cm}C{2.4cm}C{0.8cm}C{2.4cm}C{0.8cm}C{0.8cm}C{0.8cm}C{2.4cm}C{0.8cm}C{0.8cm}@{}}
\hline

    G327.6+14.6 & SNR   & 2.2   & Winkler et al.\ 2003 & 0.68$\pm$0.01 & Uchida et al.\ 2013 & A     & 0.68$\pm$0.10 & 1.6$\pm$0.1 & Foight et al.\ 2016 & A     & 1.6$\pm$0.1 & 0.32$\pm$0.1 & Schweizer \& Middleditch 1980 & A     & 0.32$\pm$0.1 \\
          &       &       &       & 0.42$\pm$0.08 & Broersen et al.\ 2013 & A     &       &       &       &       &       &       &       &       &  \\
          &       &       &       & 0.90$\pm$0.05 & Bamba et al.\ 2003 & A     &       &       &       &       &       &       &       &       &  \\
          &       &       &       & 0.56$\pm$0.06 & Allen et al.\ 2001 & A     &       &       &       &       &       &       &       &       &  \\
          &       &       &       &       &       &       &       &       &       &       &       &       &       &       &  \\
    G332.4-0.4 & SNR   & 3.1   & Caswell et al.\ 1975 & 7     & Gotthelf et al.\ 1997 & A     & 7$\pm$0.7 & 8.7$\pm$3.5 & Foight et al.\ 2016 & A     & 8.7$\pm$3.5 & 4.5   & Leibowitz \& Danziger 1983 & A     & 4.7$\pm$0.7 \\
          &       &       &       &       &       & A     &       &       &       &       &       & $\sim$5.0 & Ruiz 1983 & A     &  \\
          &       &       &       &       &       &       &       &       &       &       &       & 4.7$\pm$0.9 & Oliva et al.\ 1989 & A     &  \\
\hline
 \end{supertabular}
\end{center}
\end{minipage}
}
\end{figure}
%\end{table*}
\end{landscape}

\begin{landscape}
%\begin{table*}
\begin{figure}
%\caption{Hydrogen column density $\NH$ and distance $D$ compiled from the literature.}
{\tiny
\begin{minipage}[b]{1.0\textwidth}
\begin{center}
\begin{supertabular}{@{}C{1.2cm}C{0.25cm}C{0.25cm}C{2.4cm}C{0.8cm}C{2.4cm}C{0.8cm}C{2.4cm}C{0.8cm}C{2.4cm}C{0.8cm}C{0.8cm}C{0.8cm}C{2.4cm}C{0.8cm}C{0.8cm}@{}}
\hline

    4U 1118-60 & HMXB  & 9.0   & Liu et al.\ 2006 & 11.2$\pm$0.1 & Naik et al.\ 2011a & A     & 11.0$\pm$0.1 &       &       &       &       & 4.3   & van der Meer et al.\ 2007 & A     & 4.0$\pm$0.5 \\
          &       &       &       & 9.5$\pm$0.3 & Wojdowski et al.\ 2001 & A     &       &       &       &       &       &   &   &      &  \\
          &       &       &       &       &       &       &       &       &       &       &       & 2.8   & Coe \& Payne 1985 & R     &  \\
          &       &       &       &       &       &       &       &       &       &       &       & 3.7$\pm$0.6 & Janot-Pacheco et al.\ 1981 & A     &  \\
          &       &       &       &       &       &       &       &       &       &       &       & 4.2   & Hutchings et al.\ 1979 & A     &  \\
          &       &       &       &       &       &       &       &       &       &       &       & 3.3   & Genderen 1975 & A     &  \\
          &       &       &       &       &       &       &       &       &       &       &       & 4.2   & Mauder 1975 & A     &  \\
          &       &       &       &       &       &       &       &       &       &       &       & 4.4   & Krzeminski 1974 & A     &  \\
\hline
 \end{supertabular}
\end{center}
\end{minipage}
}
\end{figure}
%\end{table*}
\end{landscape}

\begin{landscape}
%\begin{table*}
\begin{figure}
%\caption{Hydrogen column density $\NH$ and distance $D$ compiled from the literature.}
{\tiny
\begin{minipage}[b]{1.0\textwidth}
\begin{center}
\begin{supertabular}{@{}C{1.2cm}C{0.25cm}C{0.25cm}C{2.4cm}C{0.8cm}C{2.4cm}C{0.8cm}C{2.4cm}C{0.8cm}C{2.4cm}C{0.8cm}C{0.8cm}C{0.8cm}C{2.4cm}C{0.8cm}C{0.8cm}@{}}
\hline

    4U 1820-30 & LMXB  & 7.6   & Liu et al.\ 2007 & 2.9$\pm$0.1 & Valencic \& Smith 2015 & A     & 2.7$\pm$0.1 & 2.5$\pm$0.3 & Guver et al.\ 2010b & A     & 2.6$\pm$0.3 & 0.9$\pm$0.1 & Harris 2010 and 1996 & A     & 0.9$\pm$0.1 \\
          &       &       &       & 2.0   & Yao \& Wang 2006 & A     &       & 1.9$\pm$0.1 & Mondal et al.\ 2016 & R     &       & 1.1   & Valenti et al.\ 2004 & A     &  \\
          &       &       &       &       &       &       &       & 2.7$\pm$0.4 & Juett et al.\ 2004 & A     &       & 1.0   & Heasley et al.\ 2000 & A     &  \\
          &       &       &       &       &       &       &       &       &       &       &       & 0.9   & Reed et al.\ 1988 & A     &  \\
          &       &       &       &       &       &       &       &       &       &       &       & 0.8   & Zinn 1980 & A     &  \\
          &       &       &       &       &       &       &       &       &       &       &       &       &       &       &  \\
    4U 1822-37 & LMXB  & 2.5   & Liu et al.\ 2007 & 1.2$\pm$0.2 & Iaria et al.\ 2001 & A     & 1.2$\pm$0.2 &       &       &       &       & 0.4$\pm$0.4 & Mason et al.\ 1982 & A     & 0.5$\pm$0.3 \\
          &       &       &       & 0.8$\pm$0.7 & Heinz \& Nowak 2001 & A     &       &       &       &       &       & 0.6$\pm$0.3 & Mason \& Cordova 1982 & A     &  \\
\hline
 \end{supertabular}
\end{center}
\end{minipage}
}
\end{figure}
%\end{table*}
\end{landscape}

\begin{landscape}
%\begin{table*}
\begin{figure}
%\caption{Hydrogen column density $\NH$ and distance $D$ compiled from the literature.}
{\tiny
\begin{minipage}[b]{1.0\textwidth}
\begin{center}
\begin{supertabular}{@{}C{1.2cm}C{0.25cm}C{0.25cm}C{2.4cm}C{0.8cm}C{2.4cm}C{0.8cm}C{2.4cm}C{0.8cm}C{2.4cm}C{0.8cm}C{0.8cm}C{0.8cm}C{2.4cm}C{0.8cm}C{0.8cm}@{}}
\hline
    GRO J1008-57 & HMXB  & 5.0   & Liu et al.\ 2006 & 9.6$\pm$0.5 & Yamamoto et al.\ 2014 & A     & 12.7$\pm$0.1 &       &       &       &       & 6.1$\pm$0.2 & Coe et al.\ 1994b & A     & 6.1$\pm$0.2 \\
          &       &       &       & 12.9$\pm$0.1 & Naik et al.\ 2011b & A     &       &       &       &       &       &       &       &       &  \\
          &       &       &       &       &       &       &       &       &       &       &       &       &       &       &  \\
          &       &       &       &       &       &       &       &       &       &       &       &       &       &       &  \\
    GRO J1655-40 & LMXB  &       &       & 7.1$\pm$0.3 & Kubota et al.\ 2005 & A     & 7.4$\pm$0.3 & 7.1$\pm$0.3 & Sala et al.\ 2007 & A     & 7.4$\pm$0.2 & 3.7$\pm$0.2 & Hynes et al.\ 1998 & A     & 3.7$\pm$0.3 \\
          &       &       &       & 7.1$\pm$0.8 & Yamaoka et al.\ 2001 & A     &       & 7.7$\pm$0.2 & Diaz Trigo 2007 & A     &       & 3.5   & Bianchini et al.\ 1997 & A     &  \\
          &       &       &       & 7.6$\pm$0.2 & Gierlinski et al.\ 2001 & A     &       &       &       &       &       & 4.0   & Horne et al.\ 1996 & A     &  \\
          &       &       &       & 7.4$\pm$0.3 & Ueda et al.\ 1998 & A     &       &       &       &       &       & 3.6   & Bailyn et al.\ 1995 & A     &  \\
\hline
\end{supertabular}
\end{center}
\end{minipage}
}
\end{figure}
%\end{table*}
\end{landscape}

\begin{landscape}
%\begin{table*}
\begin{figure}
%\caption{Hydrogen column density $\NH$ and distance $D$ compiled from the literature.}
{\tiny
\begin{minipage}[b]{1.0\textwidth}
\begin{center}
\begin{supertabular}{@{}C{1.2cm}C{0.25cm}C{0.25cm}C{2.4cm}C{0.8cm}C{2.4cm}C{0.8cm}C{2.4cm}C{0.8cm}C{2.4cm}C{0.8cm}C{0.8cm}C{0.8cm}C{2.4cm}C{0.8cm}C{0.8cm}@{}}
\hline

    SAX J1748.9-2021 & HMXB  & 8.5$\pm$0.4 & Liu et al.\ 2006 & 7$\pm$1 & Cackett et al.\ 2005 & A     & 7$\pm$1 &   5.7$\pm$0.1 & Pintore et al.\ 2016    &       &       & 3.1$\pm$0.3 & Ortolani et al.\ 1994a & A     & 3.2$\pm$0.4 \\
          &       &       &       &       &       &       &       &       &       &       &       & 3.3   & Armandroff 1989 & A     &  \\
          &       &       &       &       &       &       &       &       &       &       &       & 3.6   & Harris \& van den Bergh 1974 & A     &  \\
          &       &       &       &       &       &       &       &       &       &       &       &       &       &       &  \\
    RX J0146.9+6121 & HMXB  & 2.3   & Liu et al.\ 2006 & 4.9$\pm$0.5 & La Palombara \& Mereghetti 2006 & A     & 4.9$\pm$0.5 &       &       &       &       & 2.9$\pm$0.1 & Reig et al.\ 1997 & A     & 2.9$\pm$0.1 \\
          &       &       &       & 9.4$\pm$1.2 & Haberl et al.\ 1998a & R     &       &       &       &       &       & 3.4   & Motch et al.\ 1997 & A     &  \\
          &       &       &       & 7.8$\pm$1.9 & Haberl et al.\ 1998b & R     &       &       &       &       &       & 2.9   & Coe et al.\ 1993 & A     &  \\
          &       &       &       &       &       &       &       &       &       &       &       &       &       &       &  \\
\hline
\end{supertabular}
\end{center}
$^a$: We convert the measured Balmer decrement to extinction
          using
          $\AV\approx\frac{1}{0.331}\times\log\left(\frac{\HH{\alpha}}{3
              \HH{\beta}}\right)\times0.664\times3.1$ (Mavromatakis et al. 2004a).\\
$^b$: The distances are estimated based on $\AV$
          and the extinction maps of Marshall et al.\ (2006),
          Sale et al.\ (2014) and Green et al.\ (2014).\\
\end{minipage}
}
\end{figure}
%\end{table*}
\end{landscape}

%\begin{thebibliography}{}
\bibliographystyle{mn2e}

\end{document}